\documentclass[final,5p,times,twocolumn,monochrome]{elsarticle}

\usepackage{amssymb}

\newcommand{\D}[0]{{\rm d}}
\newcommand{\img}[0]{{\rm i}}
\newcommand{\ie}[0]{\textit{i.e.}}
\newcommand{\Mod}[1]{|#1|}
\usepackage{mathtools}
\usepackage{rotating}
\usepackage{amsmath}
\usepackage{multirow}
\usepackage{pgfplots}
\usepackage{xcolor}
\usepackage[normalem]{ulem}
\usepackage[font=normalsize]{caption}
\usepackage[font=normalsize]{subcaption}
\usepackage{bm}
\usepackage{soul}
\usepackage{hyperref}
\hypersetup{
    colorlinks = true,
    citecolor = blue,
    linkcolor = red,
}
\newcommand\sizeFig{0.205}
\journal{Chaos, Solitons \& Fractals}

\begin{document}

\begin{frontmatter}



\title{Emergent synchrony in oscillator networks with adaptive arbitrary-order interactions} 

\author[label1]{Dhrubajyoti Biswas}
\ead{dhrubajyoti98@gmail.com}

\author[label1]{Arpan Banerjee}
\ead{arpan@nbrc.ac.in}
\affiliation[label1]{organization={National Brain Research Centre},
            addressline={Manesar}, 
            city={Gurgaon},
            postcode={122052}, 
            state={Haryana},
            country={India}}

\begin{abstract}
Complex physical systems are often governed by interactions that extend beyond pairwise links, underscoring the need to establish a map between interpretable system parameters and emergent synchronisation phenomena in hyper-graphs. To achieve this, the current work formulates an adaptive Kuramoto model that incorporates hyperedges of arbitrary order and explores their effects on synchronisation. By deriving the exact order parameter dynamics in the thermodynamic limit, analytical expressions governing the collective dynamics are obtained. Subsequent numerics confirm the analytical predictions, in addition to capturing qualitatively different dynamical regimes and phase transitions. Further investigations based on numerically constructed order parameter distributions demonstrate how fluctuations due to finite system size can influence the long-term system dynamics by inducing spontaneous transitions. These results provide important insights and can have diverse applications, such as designing optimal surgical procedures for drug-resistant epilepsy in the human brain and identifying the sources of rumours in social networks.
\end{abstract}




\end{frontmatter}

\section{Introduction}

\begin{figure}[h]
    \centering
    \subcaptionbox{\label{fig_hoi_network_c}}{\includegraphics[width=0.475\textwidth]{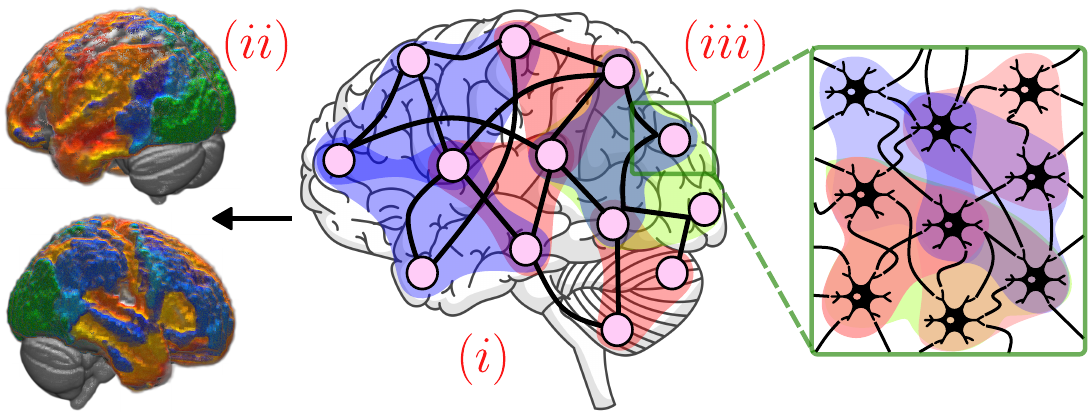}}

    \centering
    \subcaptionbox{\label{fig_strength_v_order}}{\includegraphics[width=0.235\textwidth]{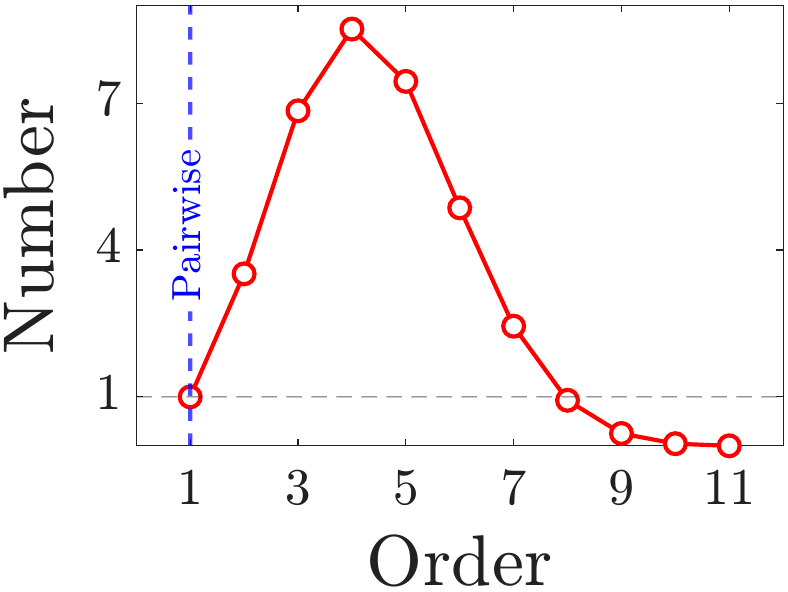}}
    \subcaptionbox{\label{fig_epilepsy_scheme}}{\includegraphics[width=0.235\textwidth]{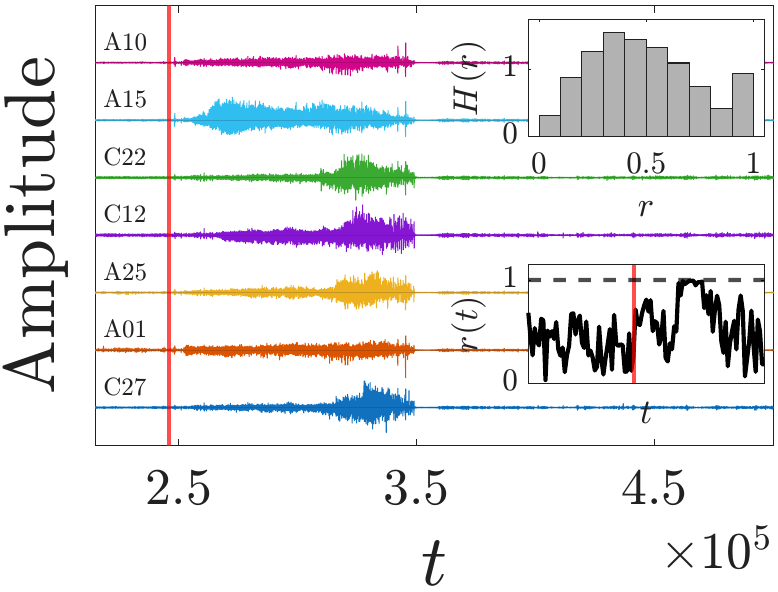}}
    \caption{(a): Schematic representation of hyperedges in the brain: In (i), pink circles indicate nodes, or Broadman areas of functional significance, whereas the lines (edges) and coloured regions (hyperedges) denote pairwise and higher-order interactions, respectively. These can appear at multiple spatial scales, such as macroscopic functional connectivity patterns from fMRI-BOLD time series and microscopic neuronal scale, shown in (ii) and (iii), respectively; (b): Number of interactions in a brain network, as a function of its order, relative to pairwise interactions; see Refs.~\cite{karmelic2022emergent,cabral2022metastable} for more details; (c): Variation of intracranial (iEEG) signals from different channels~\cite{bougou2025mesoscale,dataset}, with the histogram (top-inset) and time series (bottom-inset) of phase synchrony, highlighting its increase during epilepsy, with its onset marked with a vertical line.}
\end{figure}

\textcolor{red}{Recent literature across different scientific disciplines shows that real-life complex physical systems involve connections that extend beyond basic pairwise edges and include couplings that capture interactions between groups of three or more nodes simultaneously~\cite{hypergraphtheoryBretto2013,majhi2022dynamics,shi2022simplicial,boccaletti2023structure,PhysRevLett.131.207401}; see Fig. \ref{fig_hoi_network_c} for a sample schematic. Such interactions, mathematically represented as hyper-edges or simplicial complexes~\cite{salnikov2018simplicial}, cannot be further decomposed into their constituent pairwise interactions. Systems arising from social dynamics are an archetypal example, and include collaboration networks in scientific research~\cite{lungeanu2021team,moore2012analyzing}, group interactions in opinion spreading~\cite{wang2020social,ghahremani2023novel}, and dynamics of consensus~\cite{PhysRevE.101.032310,PhysRevE.104.064305}, among others.} Within biological systems, which are often replete with higher-order interactions, neuroscience stands out as a promising domain~\cite{giusti2016two,hindriks2025unraveling}. Higher-order connectivity is evident in the human brain~\cite{karmelic2022emergent} (also see Fig.~\ref{fig_strength_v_order}), and has been linked to avalanches and other neural dynamics~\cite{yu2011higher,millan2025spatio,biswas2025effect}, and could provide deeper insights into pathological states~\cite{PhysRevResearch.5.013074}. These also appear in models describing chemical reactions~\cite{wen2023chemical,brauch2013higher}, power grids~\cite{ghasemi2022higher,PhysRevE.109.024212}, and the evolution of ecological systems~\cite{swain2022higher,gallardo2024higher,bairey2016high,PhysRevE.111.014309}, highlighting their ubiquity across domains. Mathematically, the dynamics of such systems can be described by
\begin{align}
    \frac{\D x_i}{\D t}&=f_i(x_i)+\sum_{\mathclap{d=1;j_1,..,j_d=1}}^{D;N} A^{(d)}_{i,j_1,..,j_d} g^{(d)}(x_i,x_{j_1},..,x_{j_d}),
\label{dynsys}
\end{align}
where $N$ and $D$ denote the number of constituent nodes and the maximum order of interaction, respectively. Here, the tensor $A^{(d)}$ and the multivariate function $g^{(d)}(\cdot)$ capture the structure and mechanism of interactions, respectively, of order $d\leq D$.

Oscillations are a characteristic feature often observed in such systems, with their synchronisation representing a vital emergent phenomenon having important implications across disciplines \cite{arenas2008synchronization}. \textcolor{red}{It is particularly important in models of regular and pathological neural dynamics~\cite{ansarinasab2025synchronization,ansarinasab2025transition}}, among which a key example is epilepsy, often characterised by excess synchronous activity between different regions of the brain~\cite{osterhage2007measuring,lehnertz2009synchronization}; see Fig.~\ref{fig_epilepsy_scheme} for sample schematics.
\textcolor{red}{To this end, the Kuramoto model~\cite{RevModPhys.77.137,gupta2014kuramoto} serves as a paradigmatic framework for quantitatively exploring synchronisation in a systematic manner. In addition, to capture physically realistic complex systems, it has also been extensively modified to include features such as complex connectivity patterns~\cite{jain2023composed,chen2023critical,peron2019onset,tong2018exponential}, symmetry-breaking mechanisms~\cite{PhysRevE.102.012206,manoranjani2023generalization,biswas2024symmetry}, time-delayed interactions~\cite{yeung1999time,wu2018dynamics,ross2021dynamics,skardal2022tiered,budzinski2023analytical}, phase frustration~\cite{buzanello2022matrix,PhysRevE.108.034208,moyal2024rotating,mondal2025enhancing}, and higher-order couplings~\cite{xu2021spectrum,costa2024bifurcations,costa2025exact,rajwani2023tiered,carballosa2023cluster,sabhahit2024prolonged,skardal2020higher}, as well as different combinations of each of these}. The Kuramoto model finds applications in a myriad of different domains~\cite{guo2021overviews,wang2021dynamic,blanter2016kuramoto,fioriti2008stability}, and is of particular importance in neuroscience, where it provides a tractable mathematical framework to capture complex neuro-oscillatory phenomena~\cite{odor2019critical,heggli2019kuramoto,PhysRevLett.120.244101,odor2021effect,pathak2022biophysical,ranjan2024propagation,PhysRevE.111.044310}. \textcolor{red}{Additionally, it has also been extended to the so-called ``swarmalator'' model~\cite{o2017oscillators,PhysRevE.111.014209,ansarinasab2024spatial}, which introduces additional state variables to capture the dynamics of agents moving through space and interacting through a combination of both phase and spatial interactions, thereby demonstrating a rich repertoire of dynamical states.}

To gain an understanding of synchronisation phenomena, this article investigates an adaptive Kuramoto model that incorporates arbitrary-order interactions of \textcolor{red}{both symmetry-preserving and symmetry-breaking types} to explore the effects of higher-order interactions and their interplay with other coupling modalities. The dynamics of the system are assumed to be governed by equations of the form
\begin{multline}
    \frac{\D \phi_i}{\D t}=\omega_i+\frac{f_1(r_1) \epsilon_1}{N}\sum_{j_1=1}^N \sin\Big(\phi_{j_1}-\phi_i-\delta_1\Big)+\\\sum_{d=2}^{d=D}\left[ \frac{f_d(r_1)\epsilon_{d}}{N^d} \sum_{\mathclap{j_1,..,j_d=1}}^N \sin\left(d\phi_{j_d}-\sum_{k=1}^{d-1}\phi_{j_k}-\phi_i-\delta_{d}\right)+\right.\\ \left. \frac{f_d^{\prime}(r_1)\epsilon^{\prime}_{d}}{N^d}  \sum_{\mathclap{j_1,..,j_d=1}}^N\sin\left(d\phi_{j_d}+\sum_{k=1}^{d-1}\phi_{j_k}-\phi_i-\delta^{\prime}_{d}\right)\right],
\label{actual_eqn}
\end{multline}
\textcolor{red}{where $\phi_i(t)$ and $\omega_i$ represent the instantaneous phase and natural frequency, respectively, of each oscillator $i$. Furthermore, 
\begin{equation}
    z_{k} = r_k e^{\img \psi_k}=\frac{1}{N}\sum_{j=1}^N e^{\img k \phi_j},\quad k\geq 1,
\label{order-param-defn}
\end{equation}
are the complex Kuramoto-Daido order parameters~\cite{clusella2020irregular} that quantify the collective dynamics of the systems. In particular, for a nearly incoherent state, $r_1$ tends to zero, whereas for a highly synchronous state, $r_1$ approaches unity. In this model, the couplings between the individual phases are captured through the second, third, and fourth terms in the RHS of Eq.~\eqref{actual_eqn}, which represent the pairwise and the two separate types of higher-order interactions, respectively.  Evidently, the system is globally coupled, \ie, every node interacts with every other node in an identical manner, across both pairwise and higher-order links.}

\textcolor{red}{The interaction terms are strongly influenced by the functions $f_d(\cdot)$ and $f_d^{\prime}(\cdot)$, the inclusion of which captures adaptivity, a fundamental feature of many natural systems~\cite{sawicki2023perspectives}, that allows modulating the characteristic properties in response to the current collective state of the overall system}. It is seen across bio-physical, socio-economic, and engineering systems, and is particularly prevalent in the study of neural activity; for example, spike-frequency adaptation models the reduction in neural firing arising from the fatigue of continuous spiking activity~\cite{PhysRevE.107.024311}. In the context of synchronisation, this allows the modulation of the global coupling strengths (\ie, $\epsilon_d$ and $\epsilon_d^{\prime}$) in response to the extent of synchronisation in the system~\cite{jin2023synchronization}, quantified through $r_1$, thereby allowing the system to reach a desired state of synchronicity. For example, a universal and well-investigated form of the adaptation function is a power-law, where the exponent controls the nature of the feedback and the resulting dynamics~\cite{PhysRevE.75.017201,PhysRevE.102.012219,biswas2024effect}, \textcolor{red}{a variation of which is investigated in later sections. Furthermore, the coupling terms also comprise of phase lags in each order of interaction (\ie, $\delta_d$ and $\delta_d^{\prime}$), the primary effect of which is to modify the common frequency to some desired value~\cite{lohe2015synchronization}, whereas both symmetry-preserving and symmetry-breaking higher-order interactions terms are introduced to capture non-trivial generalizations of the basic Kuramoto model, such as the Winfree model~\cite{manoranjani2023generalization,PhysRevE.96.042208}.}

The rest of the article is arranged as follows: Sec.~\ref{anal} outlines the dimensionality reduction of the proposed system, which provides analytical insights into the overall dynamics. In Sec.~\ref{num-res}, these results are validated against numerical simulations in Sec.~\ref{char-dyn}, highlighting different dynamical regimes, with subsequent subsections showcasing the relative effect of interactions (Sec.~\ref{rel-effect}) and the effect of fluctuations in finite-sized networks (Sec.~\ref{fin-size}). The paper ends with a summary of the key results, limitations, and future research directions in Sec.~\ref{conc}.

\section{Analytical Insights}
\label{anal}

\textcolor{red}{
To obtain analytical insights into the dynamics of the proposed system, Eq.~\eqref{actual_eqn} is first rewritten, using the definition of the order parameter $z_k$ in Eq.~\eqref{order-param-defn}, in the form
\begin{multline}
    \frac{\D \phi_i}{\D t}=\omega_i+\frac{f_1(r_1) \epsilon_1}{2\img}\left(z_1e^{-\img (\delta_1+\phi_i)}-z_1^{*}e^{\img (\delta_1+\phi_i)}\right)+\\
    \sum_{d=2}^D \left[\frac{f_d(r_1) \epsilon_d}{2\img}\left(z_d(z_1^{*})^{d-1}e^{-\img(\delta_d+\phi_i)}-z^{*}_d z_1^{d-1}e^{\img(\delta_d+\phi_i)}\right)\right.+\\\left.\frac{f^{\prime}_d(r_1)\epsilon_d^{\prime}}{2\img}\left(z_d z_1^{d-1}e^{-\img (\delta^{\prime}_d+\phi_i)}-z_d^{*}(z_1^*)^{d-1}e^{\img (\delta^{\prime}_d+\phi_i)}\right)\right],
\label{phidoteqn}
\end{multline}
where $*$ represents complex conjugation.
For an infinitely large number of oscillators, an assumption which is relaxed in the later sections, the state of the system in Eq.~\eqref{actual_eqn} can be described by a joint probability density function $F(\phi,\omega;t)$, which satisfies a continuity equation of the form 
\begin{equation}
    \frac{\partial F}{\partial t}+\frac{\partial (F\dot{\phi})}{\partial \phi}=0
\label{continuity}
\end{equation}
and the condition
\begin{equation}
    \int_{0}^{2\pi} F(\phi,\omega;t)\ \D\phi=g(\omega),
\end{equation}
where $\dot{\phi}$ is given by Eq.~\eqref{phidoteqn}. Since $F(\phi+2\pi,\omega;t)=F(\phi,\omega;t)$, it can also be expanded as a series of the form
\begin{equation}
    F(\phi,\omega;t)=\frac{g(\omega)}{2\pi}\left[1+\sum_{n=1}^{\infty}\left(F_ne^{\img n \phi}+F^*_ne^{-\img n \phi}\right)\right],
\label{fourier}
\end{equation}
where $F_n\equiv F_n(\omega,t)$ is the $n^{\rm th}$ Fourier coefficient.
By using the Ott-Antonsen ansatz~\cite{ott2008low,ott2009long}, the Fourier modes are assumed to be of the form $F_n(\omega,t)=v^n(\omega,t)$, with the restriction $\Mod{v(\omega,t)}<1$ so that the sum in Eq.~\eqref{fourier} converges. This implies that Eq.~\eqref{continuity} can be re-written in terms of $v(\omega,t)$ as
\begin{multline}
    \frac{\partial v}{\partial t}+i\omega v + \frac{f_1(r_1)\epsilon_1}{2}\left(z_1e^{-\img \delta_1}v^2-z_1^*e^{\img \delta_1}\right)+\\
    \sum_{d=2}^D \left[ \frac{f_d(r_1) \epsilon_d}{2}\left(z_d(z_1^*)^{d-1}e^{-\img \delta_d}v^2-z_d^* z_1^{d-1}e^{\img \delta_d}\right)\right.+\\\left.\frac{f^{\prime}_d(r_1) \epsilon_d^{\prime}}{2}\left(z_d z_1^{d-1}e^{-\img \delta^{\prime}_d}v^2-z_d^*(z_1^*)^{d-1}e^{\img \delta^{\prime}_d}\right)\right]=0.
\label{v-eqn}
\end{multline}
Assuming that the natural frequencies $\omega_i$ are sampled from a Cauchy distribution with mean $\omega_0$ and scale factor $\Delta$, the summation in Eq.~\eqref{order-param-defn} can be replaced with an integral, which can then be evaluated to obtain
\begin{equation}
    z_d=\int_{0}^{2\pi}\int_{-\infty}^{\infty} e^{\img d \phi}F(\phi,\omega;t)\ \D \omega\ \D \phi=\Big[v^*(\omega_0-i\Delta,t)\Big]^d
    \label{zequal}.
\end{equation}
}
Upon by substituting Eq.~\eqref{zequal} into Eq.~\eqref{v-eqn} and re-labelling $z_1$ as $z$ for the sake of simpler notation, Eq.~\eqref{v-eqn} takes the form
\begin{multline}
    \frac{\D z}{\D t}=\img(\omega_0+\img \Delta)z+\frac{f_1(r)\epsilon_1}{2}\left(ze^{-\img \delta_1}-z \Mod{z}^2e^{\img \delta_1}\right)+\\
    \sum_{d=2}^D  \left[\frac{f_d(r) \epsilon_d}{2}\left(\Mod{z}^{2(d-1)}ze^{-\img \delta_d}-\Mod{z}^{2d}ze^{\img \delta_d}\right)\right.+\\ \left.\frac{f^{\prime}_d(r) \epsilon^{\prime}_d}{2}\left(z^{2d-1}e^{-\img \delta^{\prime}_d}-\Mod{z}^4(z^*)^{2d-3} e^{\img \delta^{\prime}_d}\right)\right],
\label{z-eqn}
\end{multline}
yielding a two-dimensional coupled system, the analysis of which is simpler compared to Eq.~\eqref{actual_eqn}. \textcolor{red}{First, the stability of the incoherent state of the system can be obtained by linearizing Eq.~\eqref{z-eqn} about $z(t)=0$, leading to the equation
\begin{equation}
    \frac{\D z}{\D t}=\img(\omega_0+\img \Delta)z+\frac{z}{2}f_1(0)\epsilon_1 e^{-\img \delta_1}.
\label{lin-z}
\end{equation}
This implies that $z=0$ is stable only for
\begin{equation}
    \epsilon_1\leq \epsilon_1^c=\frac{2\Delta}{f_1(0)\cos(\delta_1)},
\label{forward-point}
\end{equation}
which corresponds to the critical point beyond which the system undergoes a phase transition from an asynchronous state to a synchronous state. Alternatively, Eq.~\eqref{z-eqn} can also be rewritten as
\begin{align}
\begin{split}
    \frac{\D r}{\D t}+\img r\frac{\D \psi}{\D t}&=\img(\omega_0+\img \Delta)r+\frac{r f_1(r)\epsilon_1}{2}H(r,\delta_1)+\\ &\sum_{d=2}^D r^{2d-1} \left[\frac{f_d(r) \epsilon_d}{2}H(r,\delta_d)+\frac{f^{\prime}_d(r) \epsilon_d^{\prime}}{2}H(r,\chi^{\prime}_d)\right],
\label{combined-eqn}
\end{split}
\end{align}
where $H(x,y)=(1-x^2)\cos(y)-\img (1+x^2)\sin(y)$ and $\chi^{\prime}_d=\delta^{\prime}_d-2(d-1)\psi$.} Therefore, the dynamics of the synchronous states (\ie, where $r>0$) are governed by coupled differential equations of the form
\begin{multline}
    \frac{\D r}{\D t}=R(r,\psi)=-\Delta r+r(1-r^2)\left[\frac{f_1(r)\epsilon_1}{2}\cos(\delta_1)\right.+\\\left.\sum_{d=2}^D r^{2(d-1)}  \left(\frac{f_d(r) \epsilon_d}{2}\cos(\delta_d)+\frac{f^{\prime}_d(r) \epsilon_d^{\prime}}{2}\cos(\chi^{\prime}_d)\right)\right],
\label{reqn}
\end{multline}
and
\begin{multline}
    \frac{\D \psi}{\D t}=\Psi(r,\psi)=\omega_0-(1+r^2)\left[\frac{f_1(r)\epsilon_1}{2}\sin(\delta_1)\right.+\\\left.
    \sum_{d=2}^D r^{2(d-1)}  \left(\frac{f_d(r) \epsilon_d}{2} \sin(\delta_d)+\frac{f^{\prime}_d(r) \epsilon_d^{\prime}}{2}\sin(\chi^{\prime}_d)\right)\right],
\label{psieqn}
\end{multline}
\textcolor{red}{which can be obtained by separating the real and imaginary parts of Eq.~\eqref{combined-eqn}}. The steady state solutions of the above two equations (say, $(r_s,\psi_s)$) provide insights into the system dynamics. \textcolor{red}{These can be obtained as solutions of the algebraic equations $R(r_s,\psi_s)=0$ and $\Psi(r_s,\psi_s)=0$, which can be expressed as
\begin{equation}
    \left(\frac{\Delta}{1-r^2_s}\right)-\sum_{n=1}^D \sigma_n(r_s)\cos(\delta_n)=\sum_{m=2}^D\sigma_m^{\prime}(r_s)\cos(\chi_{m,s}^{\prime}),
\label{ss1}
\end{equation}
and
\begin{equation}
    \left(\frac{\omega_0}{1+r^2_s}\right)-\sum_{n=1}^D \sigma_n(r_s)\sin(\delta_n)=\sum_{m=2}^D\sigma_m^{\prime}(r_s)\sin(\chi_{m,s}^{\prime}),
\label{ss2}
\end{equation}
where $\sigma_n(r)=f_n(r)\epsilon_nr^{2(n-1)}/2$, $\sigma^{\prime}_n(r)=f^{\prime}_n(r)\epsilon^{\prime}_nr^{2(n-1)}/2$, and $\chi^{\prime}_{m,s}$ denotes $\chi^{\prime}_{m}$ evaluated at $\psi={\psi_s}$. Obtaining their stability entails calculating the Jacobian $J$ of Eqs.~\eqref{reqn} and \eqref{psieqn}, the analytical expressions of its elements $J_{ij}$ are listed in Eqs.~\eqref{J11}-\eqref{J22} of \ref{app-jaco}}. 

\textcolor{red}{Therefore, taken together, the analytical results presented in this section provide a theory of synchronization in systems with arbitrary-order interactions, while also accounting for the effects of different adaptive mechanisms, phase lags, and generalized forms of interactions that break rotational symmetry in the Kuramoto model. The subsequent sections are devoted to verifying these results through numerical simulations, as well as investigating other properties of the proposed system through numerical simulations.}

\section{Numerical Results}
\label{num-res}

\textcolor{red}{Although the analytical results derived in the previous section constitute a general, arbitrary-order theoretical framework, further analytical progress is feasible only under specific simplifying assumptions on the parameter values, whereas incorporating arbitrarily large orders of interactions would increase simulation times beyond reasonable limits. Therefore, to validate the obtained analytical expressions, the simplest yet representative example of Eq.~\eqref{actual_eqn} is chosen.} It corresponds to a system consisting of pairwise and triadic interactions, similar to that reported in Ref.~\cite{biswas2024symmetry}, and is described by equations of the form
\begin{align}
\begin{split}
    \frac{\D \phi_i}{\D t}=&\omega_i+\frac{f_1(r)\epsilon_1}{N}\sum_{j=1}^N\sin(\phi_{j}-\phi_i)+\\
    &\frac{f_2(r)\epsilon_2}{N^2}\sum_{j,k=1}^N\Bigg[\sin(2\phi_k-\phi_j-\phi_i)+\xi \sin(2\phi_k+\phi_j-\phi_i)\Bigg],
\label{pairwise_symmbreak}
\end{split}
\end{align}
where $\xi=\epsilon_2^{\prime}/\epsilon_2$. For the sake of simplicity, $f_d(r)$ are assumed to be of the form $f_d(r)=(1+\gamma r)^{\pm 2}$ for $d=1$ and $d=2$, respectively, whereas $\gamma$ is a constant \textcolor{red}{that generates a family of qualitatively different functional forms}; \textcolor{red}{also see Fig.~\ref{adapfuncs}.
\begin{figure}[h]
\centering
    \subcaptionbox{$d=1$}{\includegraphics[scale=0.315]{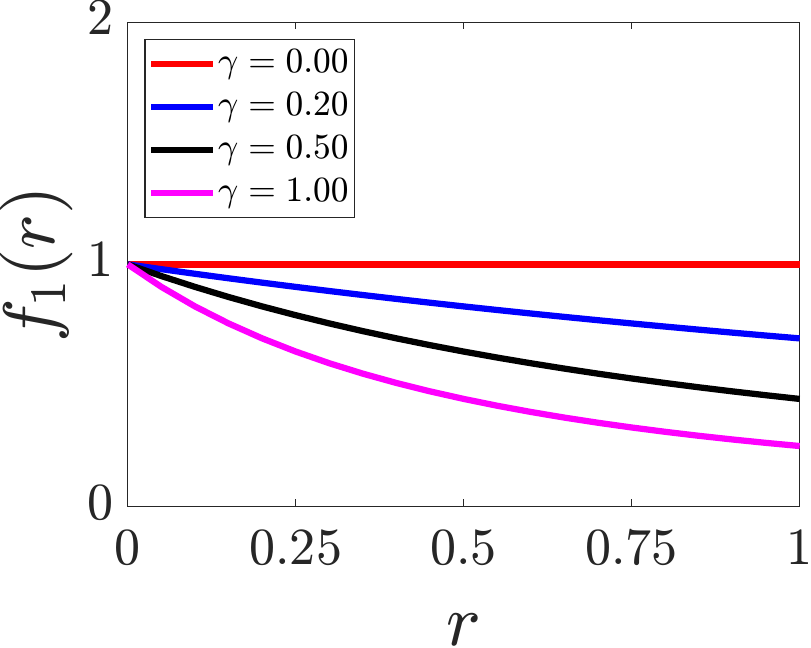}}
    \subcaptionbox{$d=2$}{\includegraphics[scale=0.315]{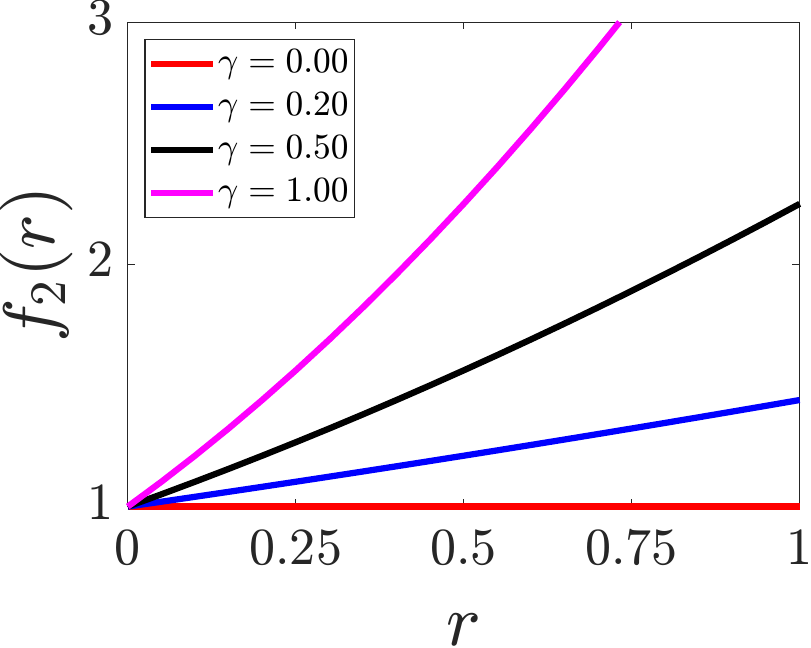}}
    \caption{\textcolor{red}{Plot of $f_d(r)$ as a function of $r\in[0,1]$ for (a) $d=1$ and (b) $d=2$, where the colours denote different values of $\gamma$.}}
    \label{adapfuncs}
\end{figure}
This choice represents the simplest possible functional form of a polynomial adaptation function, which has previously been shown to induce both continuous and explosive phase transitions~\cite{biswas2024effect}, and has been set up such that it introduces qualitatively different adaptation behaviour (\ie, increasing and decreasing) across the two orders of interactions.}
\begin{figure*}[h]
    \centering
    \subcaptionbox{$\epsilon_1=0.6$\label{fig_ts_sols_e}}{\includegraphics[scale=0.345]{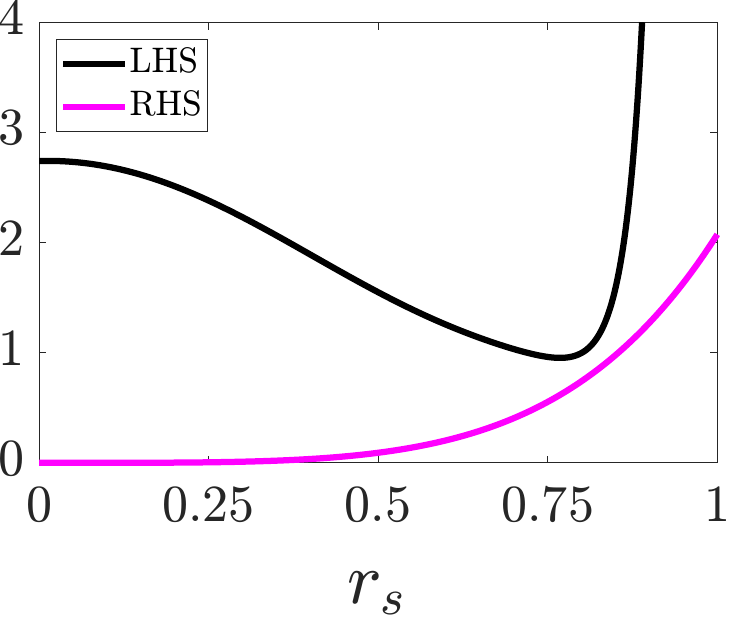}\includegraphics[scale=0.345]{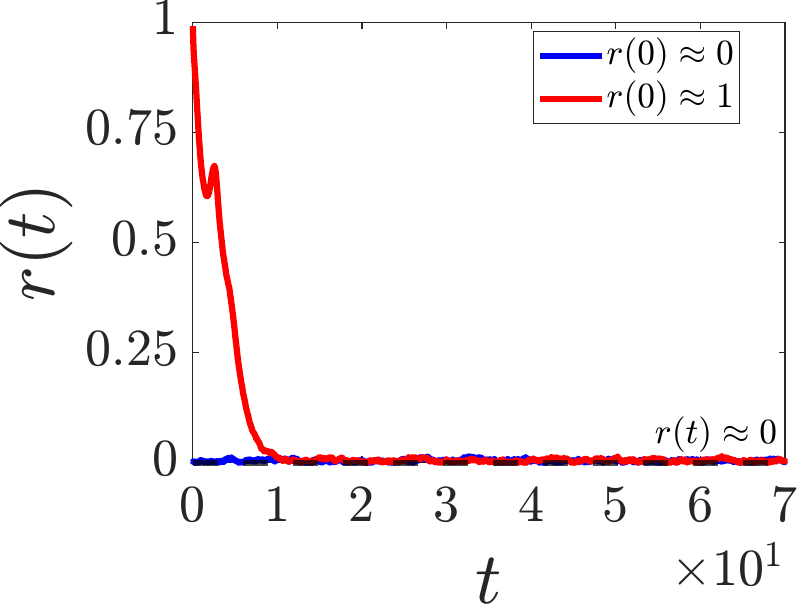}}\hfill
    \subcaptionbox{$\epsilon_1=1.3$\label{fig_ts_sols_f}}{\includegraphics[scale=0.345]{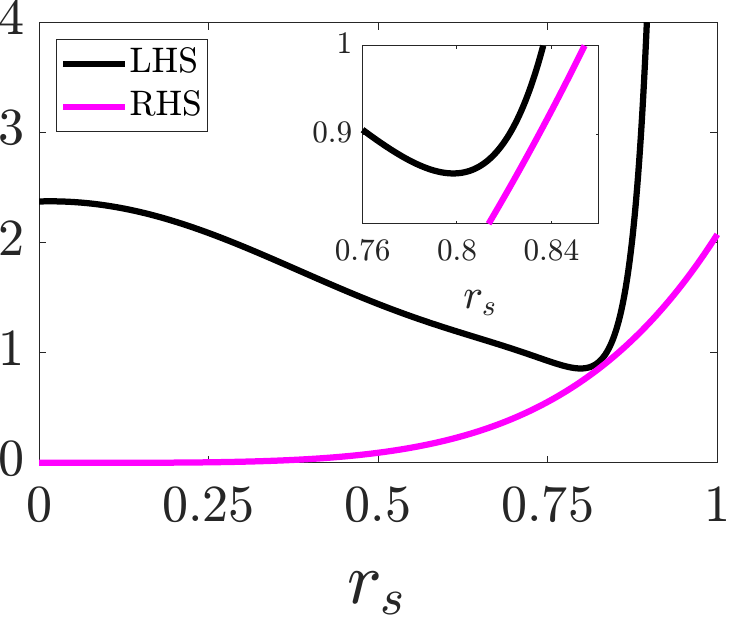}\includegraphics[scale=0.345]{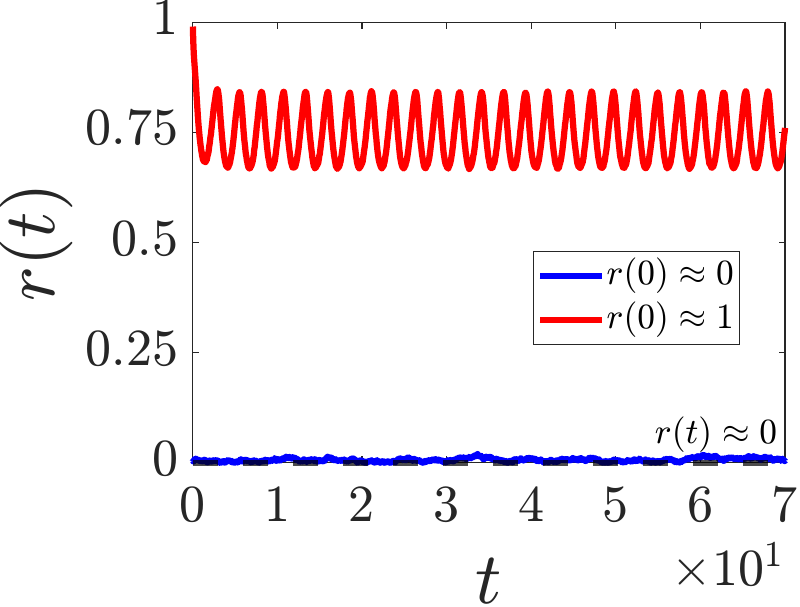}}
    
    \subcaptionbox{$\epsilon_1=1.8$\label{fig_ts_sols_g}}{\includegraphics[scale=0.345]{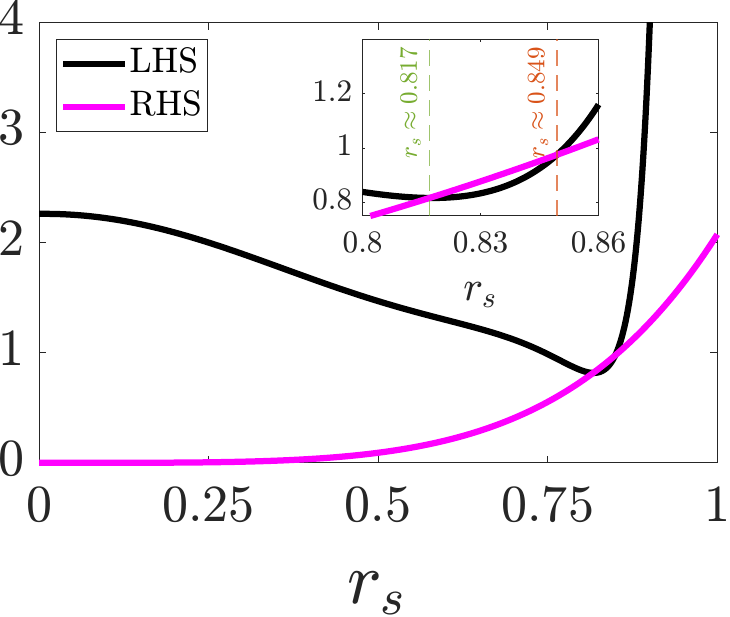}\includegraphics[scale=0.345]{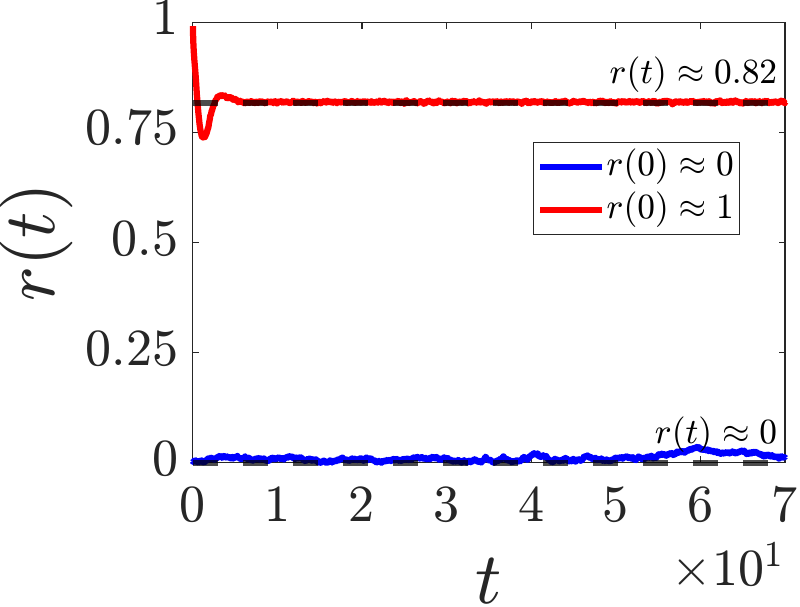}}\hfill
    \subcaptionbox{$\epsilon_1=2.6$\label{fig_ts_sols_h}}{\includegraphics[scale=0.345]{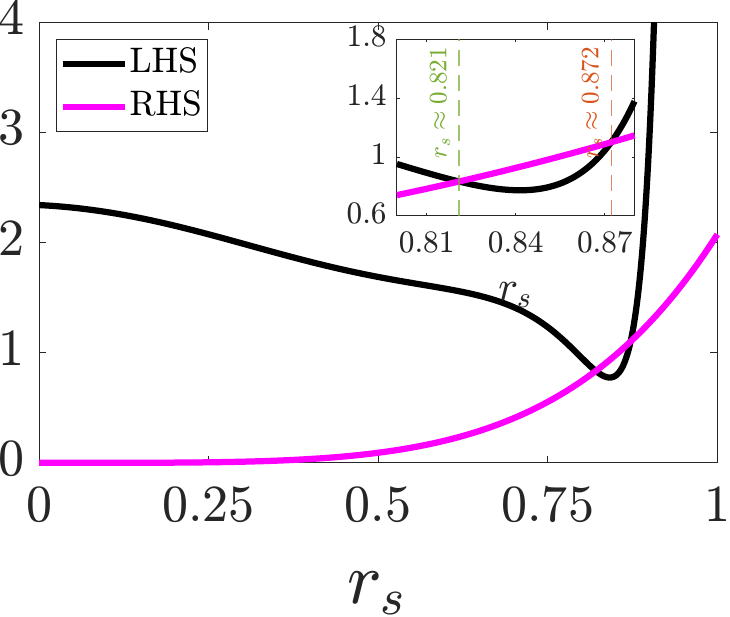}\includegraphics[scale=0.345]{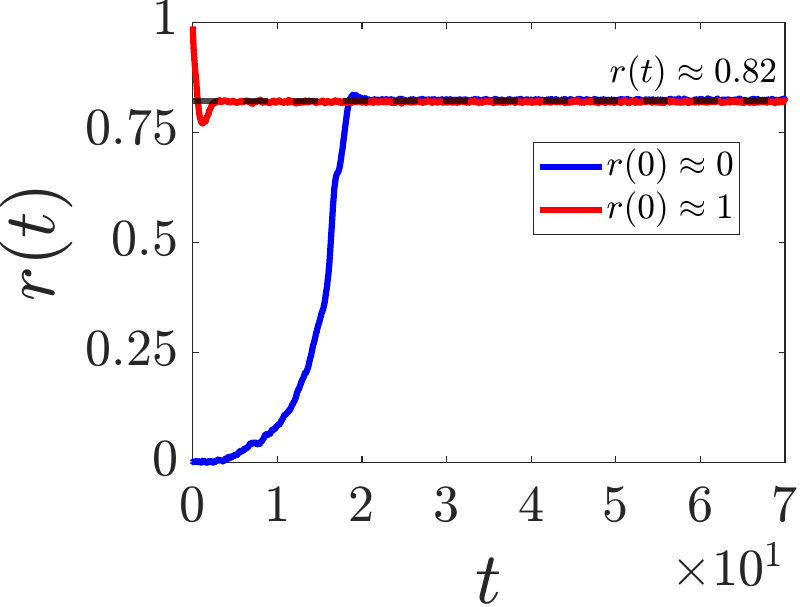}}

    \caption{(a)-(d): \emph{Left}: Graphical solution of Eq.~\eqref{final-r-eqn}. The black and magenta curves denotes the LHS and RHS, respectively, whereas the green and red broken vertical lines (in inset) denote the stable and unstable steady-state solutions respectively, as determined from the eigenvalues of the Jacobian of Eqs.~\eqref{ssh1} and \eqref{ssh2}; \emph{Right}: Temporal variation of order parameter $r(t)$. The blue and red curves denote the asynchronous and synchronous initial conditions, respectively, whereas the horizontal line denotes $r_s$. \textcolor{red}{For all figures, $\omega_0=1.5$, $\gamma=0.2$, and $\epsilon_2=5.0$}.}
    \label{fig_ts_sols}
\end{figure*}
In this case, Eqs.~\eqref{reqn} and \eqref{psieqn} take the form
\begin{align}
    \frac{\D r}{\D t}&=-\Delta r + r(1-r^2)\Big[\sigma_1(r)+ \sigma_2(r)\Big(1+\xi\cos(2\psi)\Big)\Big],\label{ssh1}\\
    \intertext{and}
    \frac{\D \psi}{\D t}&=\omega_0+\xi(1+r^2)\sigma_2(r)\sin(2\psi)\label{ssh2},
\end{align}
respectively, \textcolor{red}{where the forward phase transition point is obtained as $\epsilon_1^c=2\Delta$ from Eq.~\eqref{forward-point}}. Therefore, \textcolor{red}{the value of $r$ in equilibrium (\ie, $r_s$)} can be obtained as a solution of the algebraic equation
\begin{equation}
    \left[\frac{\Delta}{1-r_s^2}-\Big(\sigma_1(r_s)+\sigma_2(r_s)\Big)\right]^2+ \left[\frac{\omega_0}{1+r_s^2}\right]^2=\xi^2 \sigma_2^2(r_s),
\label{final-r-eqn}
\end{equation}
the stability of which is obtained from the Jacobian of Eqs.~\eqref{ssh1} and \eqref{ssh2}; see Eqs.~\eqref{j11_simple}-\eqref{j22_simple} in \ref{app-jaco}. However, finding closed-form analytical solutions is not always feasible, thus requiring the use of numerical techniques. \textcolor{red}{Furthermore, even while limited to triadic interactions, the computational load for simulating Eq.~\eqref{pairwise_symmbreak} is extensive due to the presence of nested summations in the coupling term, especially for large system sizes. This necessitates the development of alternative schemes that reduce the computational burden, an example of which is outlined in \ref{app2}}.

\subsection{Characteristic system dynamics}
\label{char-dyn}

For the rest of the work, $(\Delta,\xi)=(1.0,0.4)$ in Eq.~\eqref{pairwise_symmbreak} for the sake of simplicity. Furthermore, as an example of adaptive systems\footnote{The non-adaptive case (\ie, $\gamma=0$) has already been studied in Ref.~\cite{biswas2024symmetry}, facilitating direct comparison.}, the specific case of $\gamma=0.2$ is demonstrated in Fig.~\ref{fig_ts_sols}, for different values of $\epsilon_1$, whereas the value of \textcolor{red}{$\omega_0$ and $\epsilon_2$ is set at $1.5$ and $5.0$}, respectively.
In this scenario, for $\epsilon_1=0.6$, it is evident that Eq.~\eqref{final-r-eqn} admits no solutions. Furthermore, the asynchronous state is stable since $\epsilon_1<\epsilon_1^{c}$. Thus, the system always converges towards $r(t)\approx 0$, irrespective of initial conditions; see Fig.~\ref{fig_ts_sols_e}. However, as the value of $\epsilon_1$ is increased to $1.3$, there is a qualitative change in the dynamics. In this case, the asynchronous state remains stable, since $\epsilon_1<\epsilon_1^{c}$, and is evident from the temporal variation of $r(t)$, which converges towards $r(t)\approx 0$ when initialised in an asynchronous state. However, the system also demonstrates an unsteady synchronised state, where the order parameter shows temporal oscillations as highlighted in Fig.~\ref{fig_ts_sols_f}, \textcolor{red}{a behaviour previously observed in Ref.~\cite{biswas2024symmetry} in a similar model, and also in Ref.~\cite{biswas2022mirroring} where it emerges due to the formation of partial clusters of oscillators which instantaneously overlap and separate.}

\begin{figure*}[ht]
    \centering
    \subcaptionbox{$\gamma=0.05$\label{transition_diag_a}}{\includegraphics[scale=0.26]{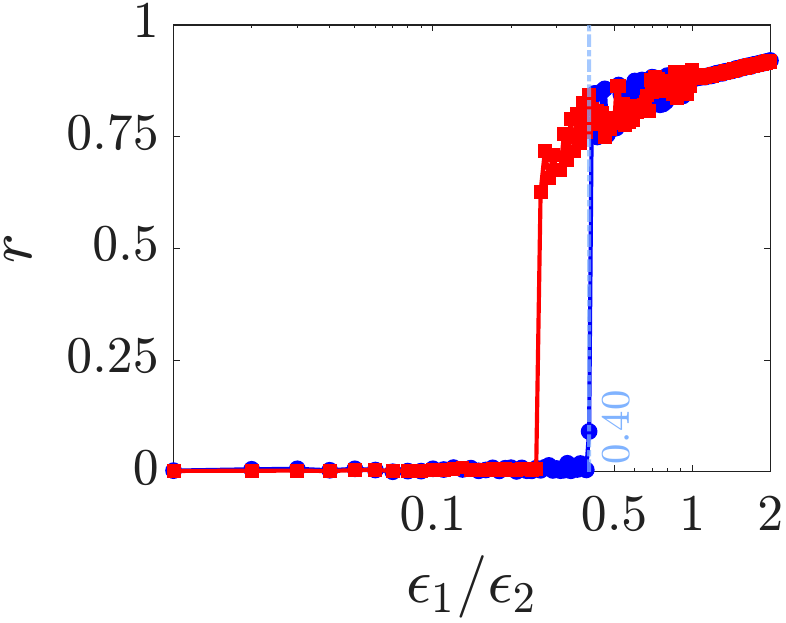}}\hfill
    \subcaptionbox{$\gamma=0.10$\label{transition_diag_b}}{\includegraphics[scale=0.26]{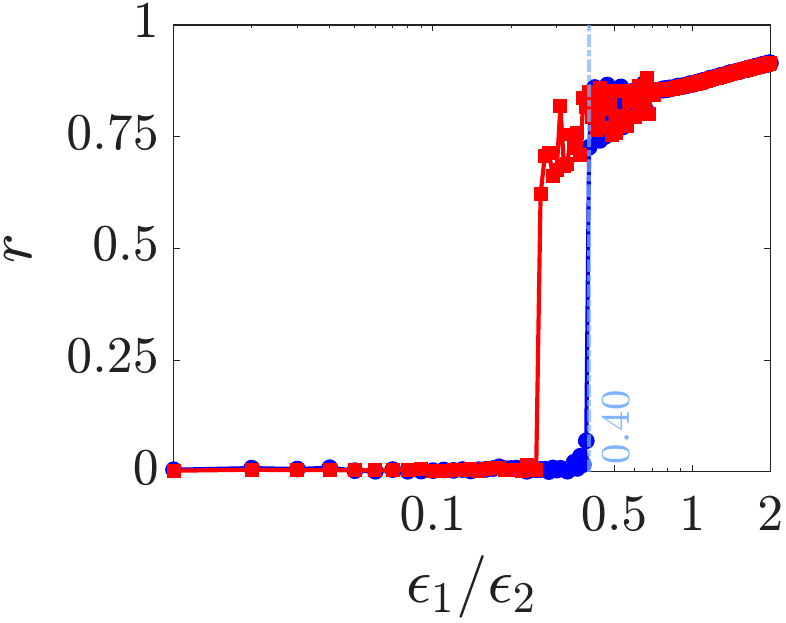}}\hfill
    \subcaptionbox{$\gamma=0.15$\label{transition_diag_c}}{\includegraphics[scale=0.26]{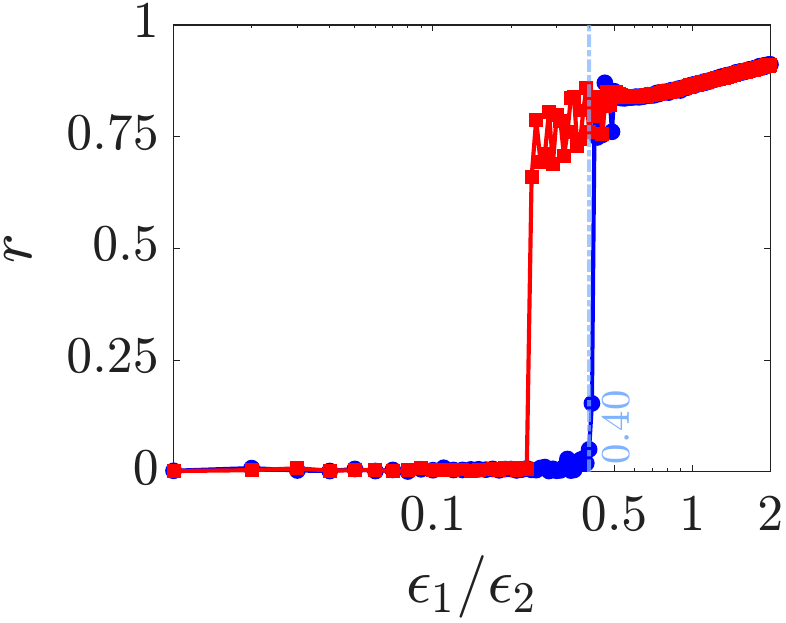}}\hfill
    \subcaptionbox{$\gamma=0.20$\label{transition_diag_d}}{\includegraphics[scale=0.26]{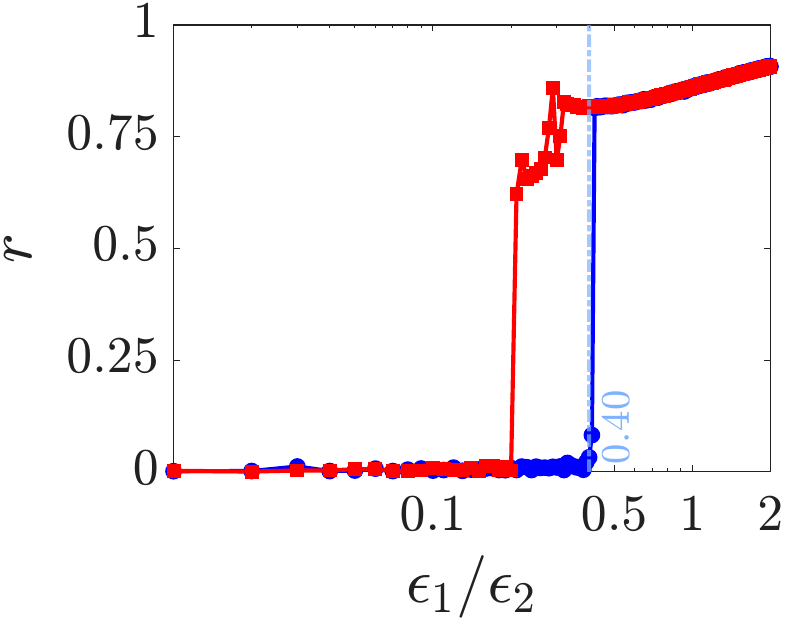}}\hfill
    \subcaptionbox{$\gamma=0.25$\label{transition_diag_e}}{\includegraphics[scale=0.26]{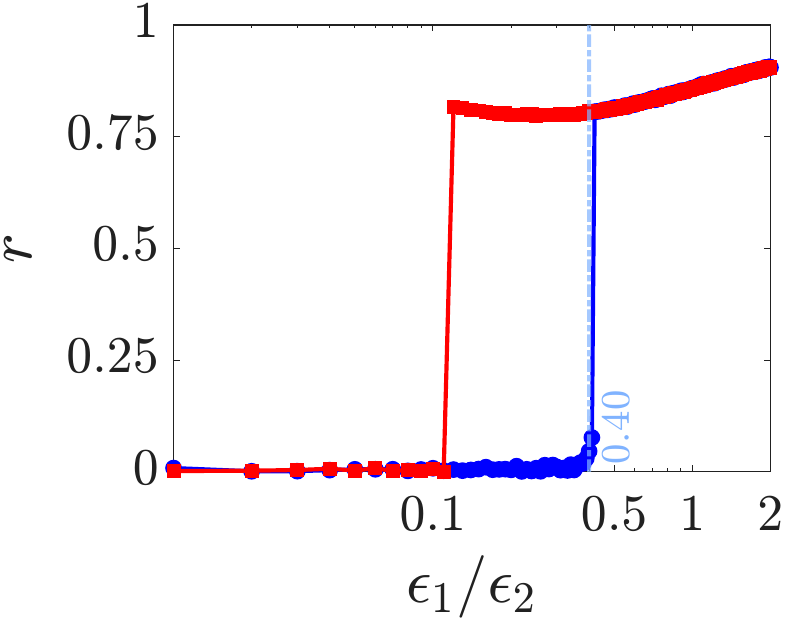}}\hfill

    \subcaptionbox{$\epsilon_2=2.0$\label{transition_diag_f}}{\includegraphics[scale=0.26]{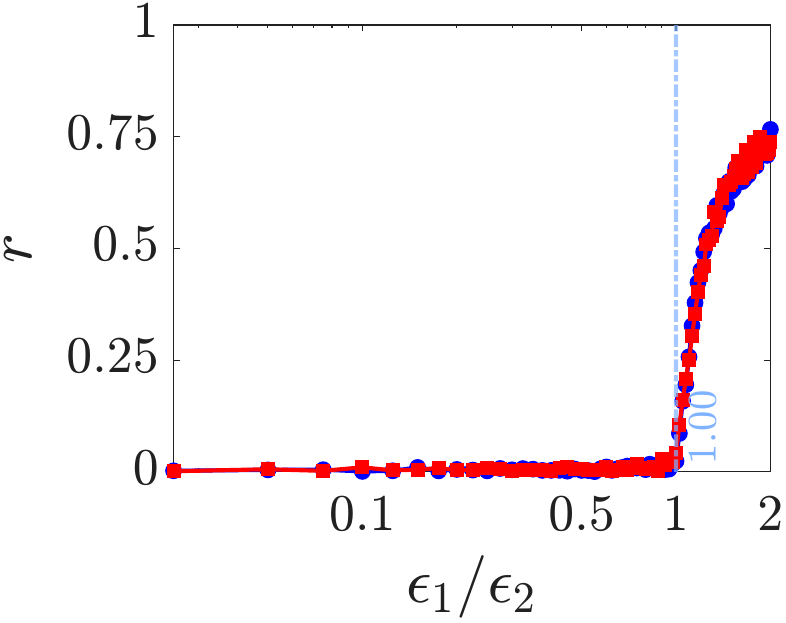}}\hfill
    \subcaptionbox{$\epsilon_2=3.0$\label{transition_diag_g}}{\includegraphics[scale=0.26]{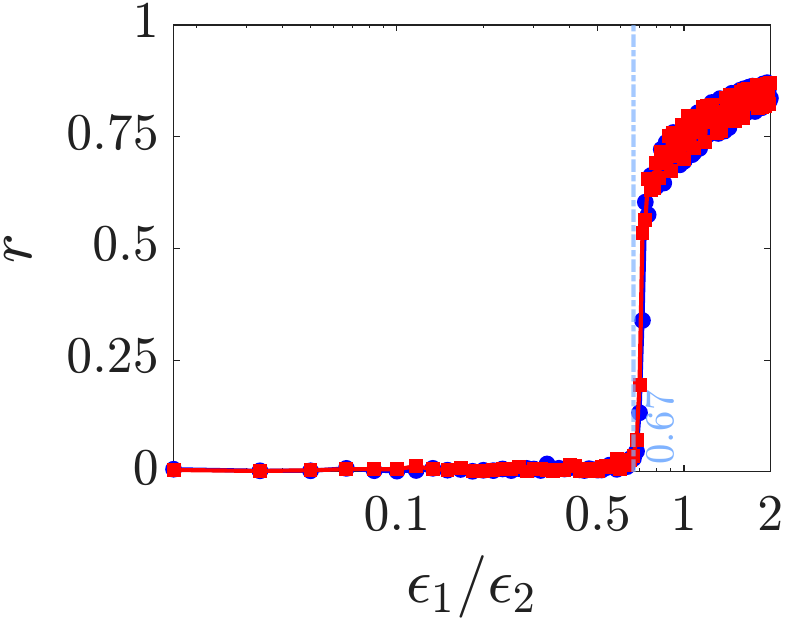}}\hfill
    \subcaptionbox{$\epsilon_2=4.0$\label{transition_diag_h}}{\includegraphics[scale=0.26]{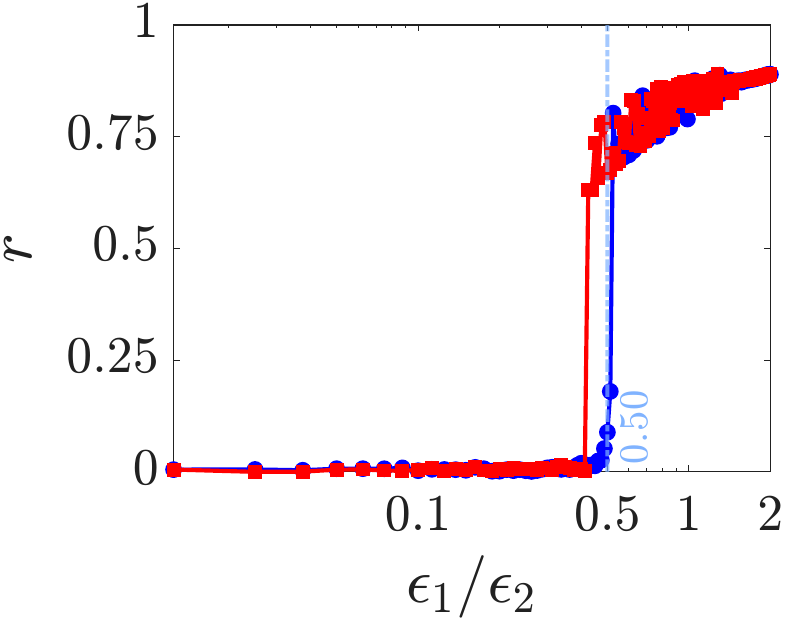}}\hfill
    \subcaptionbox{$\epsilon_2=5.0$\label{transition_diag_i}}{\includegraphics[scale=0.26]{r_e1_plot_gamma_e2_5.000_0.200_w0_1.500_fig.pdf}}\hfill
    \subcaptionbox{$\epsilon_2=6.0$\label{transition_diag_j}}{\includegraphics[scale=0.26]{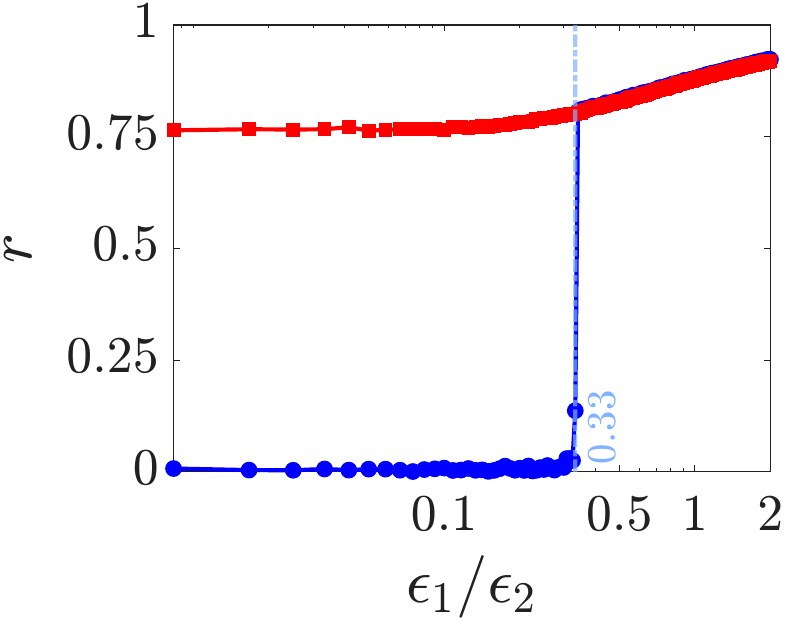}}
    \caption{(a)-(j): Variation of the order parameter $r$ as a function of $E=\epsilon_1/\epsilon_2\in[0,2]$ for different values of $\gamma$ (top-row; $\epsilon_2=5.0$), and $\epsilon_2$ (bottom-row; $\gamma=0.2$). The vertical line corresponds to the forward transition point (\ie, $E^c=\epsilon_1^c/\epsilon_2$), whereas the blue and red curves denote the forward and backward variation of $E$, respectively, \textcolor{red}{whereas $\omega_0=1.5$ across all figures}.}
    \label{transition_diag}
\end{figure*}
The observations in Fig.~\ref{fig_ts_sols_g} demonstrate that increasing $\epsilon_1$ to $1.8$ entails further change in the dynamics. In this case, Eq.~\eqref{final-r-eqn} admits two solutions $r_s=0.817$ and $r_s=0.849$. The eigenvalue analysis of these solutions demonstrates that the former is stable, whereas the latter is unstable in nature. However, the asynchronous state is also stable since $\epsilon_1<\epsilon_1^{c}$. Thus, the system is bistable, where the temporal variation of $r(t)$ reaches two different steady states depending on the initial condition. In particular, the asynchronous initial conditions continue to be asynchronous, whereas the synchronous initial condition converges to $r(t)\approx 0.82$. Finally, upon further increasing $\epsilon_1$ to $2.6$, the asynchronous state loses stability since $\epsilon_1>\epsilon_1^{c}$, as demonstrated in Fig.~\ref{fig_ts_sols_h}. However, Eq.~\eqref{final-r-eqn} continues to have two solutions, one at $r_s=0.821$, which is stable, and the other at $r_s=0.872$, which is unstable, as determined by the eigenvalue analysis. Thus, the system is now monostable, which is demonstrated by the temporal variation of $r(t)$, which converges around $r(t)\approx 0.82$ irrespective of the initial state of the system.

\subsection{Relative effect of interactions}
\label{rel-effect}
To quantify the relative effect of pairwise and higher-order interactions, the dynamics of the system are investigated via the variation in the steady state value of the order parameter as a function of the ratio $E=\epsilon_1/\epsilon_2\in[0,2]$ in Fig.~\ref{transition_diag}, representing the relative strength of the pairwise and triadic interactions.  The top and bottom rows correspond to $\epsilon_2=5.0$ and $\gamma=0.2$, respectively. In the former, the value of $\gamma$ varies between $[0.05,0.25]$ in steps of $0.05$ across the sub-figures, whereas in the latter, the value of $\epsilon_2$ varies between $[2.0,6.0]$ in steps of $1.0$.

In Figs.~\ref{transition_diag_a}-\ref{transition_diag_e}, upon increasing $\gamma$, the transition to synchrony upon forward variation of $E$ is observed to be explosive, which is characterised by a sudden increase in $r$ at the critical point $E \approx 0.4$. In contrast, the backward transition point, captured by the sharp drop in $r$ upon reducing $E$, reduces with increasing values of $\gamma$. However, the post-transition state upon increasing $E$, as well as the route to desynchronization upon reducing $E$, varies for different values of $\gamma$. 
In particular, for $\gamma=0.05$, the system demonstrates an unsteady synchronous state characterised by an oscillatory temporal variation of $r(t)$ immediately after the explosive transition. Furthermore, tuning $E$ beyond $1.0$ results in a steady synchronous state; see Fig.~\ref{transition_diag_a}. Subsequently, upon increasing $\gamma$ to $0.10$ and $0.15$, the region of $E$ for which the system demonstrates an unsteady synchronous state reduces, resulting in a larger parameter range for which the steady synchronous state is observed; see Figs.~\ref{transition_diag_b} and \ref{transition_diag_c}. Furthermore, upon increasing $\gamma$ to $0.20$ and $0.25$, the system exhibits only a steady synchronous state after the transition. However, a key difference is that, in the former, the system exhibits an unsteady synchronous state in a portion of the bi-stable regime before desynchronization, whereas the latter maintains the steady synchronous state en route to desynchronization; see Figs.~\ref{transition_diag_d} and \ref{transition_diag_e}. 

The curves in Figs. \ref{transition_diag_f}-\ref{transition_diag_j} demonstrate the qualitative dynamical changes upon changing $\epsilon_2$. Unlike the previous case, changing $\epsilon_2$ modifies the characteristic nature of the transition to synchrony upon varying $E$. In particular, while for $\epsilon_2=2.0$, the transition is observed to be \textcolor{red}{continuous} with a smooth increase in the steady-state value of $r$ beyond the critical point (see Fig.~\ref{transition_diag_f}), higher values of $\epsilon_2$ result in explosive synchronization transitions, in addition to the formation of a bi-stable zone; see Figs.~\ref{transition_diag_g}-\ref{transition_diag_j}. Similar to the previous case, the post-transition synchronous states can be both unsteady or steady (for example, see Figs.~\ref{transition_diag_h} and \ref{transition_diag_i}), whereas, for $\epsilon_2=6.0$, the system does not undergo any de-synchronisation transition upon reducing $E$; see Fig.~\ref{transition_diag_j}.

\subsection{Effect of finite system size}
\label{fin-size}
\begin{figure}[h]
    \centering
    \subcaptionbox{$\epsilon_1=0.6$\label{N50_distribution_a}}{\includegraphics[scale=0.32]{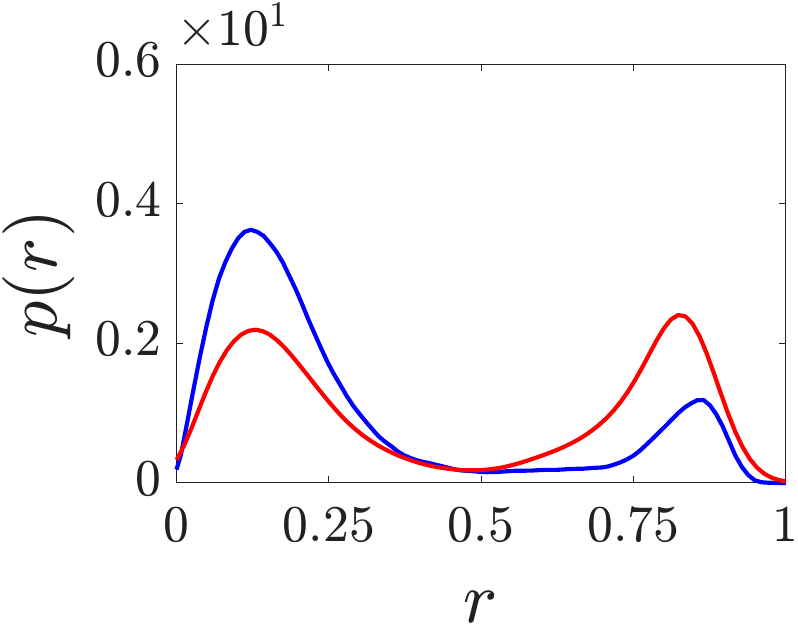}}
    \subcaptionbox{$\epsilon_1=1.3$\label{N50_distribution_b}}{\includegraphics[scale=0.32]{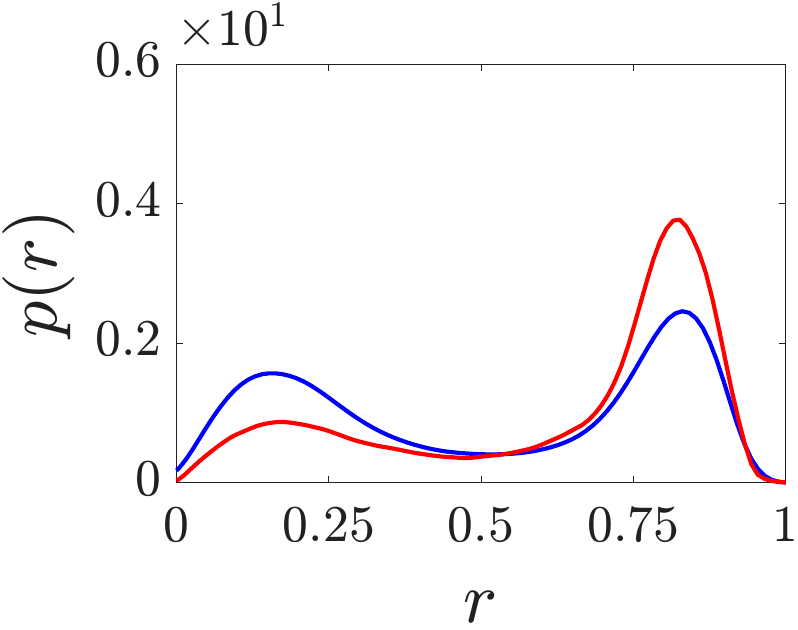}}
    \subcaptionbox{$\epsilon_1=1.8$\label{N50_distribution_c}}{\includegraphics[scale=0.32]{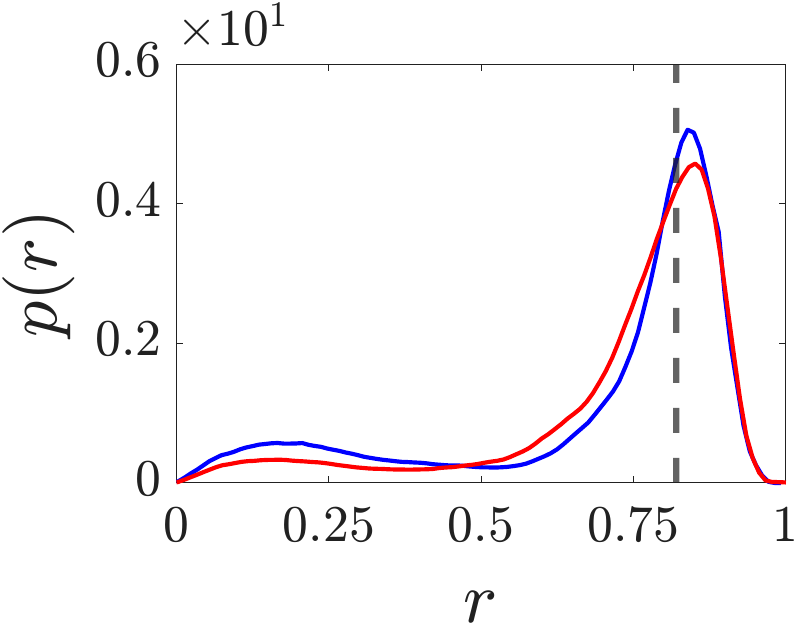}}
    \subcaptionbox{$\epsilon_1=2.6$\label{N50_distribution_d}}{\includegraphics[scale=0.32]{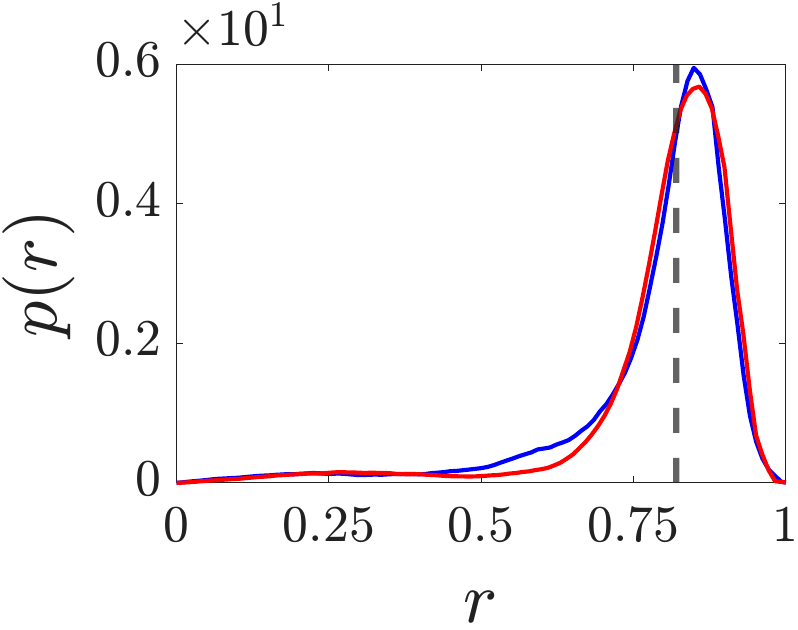}}
    
    \caption{(a)-(d): \textcolor{red}{Numerically constructed distribution of the order parameter}, \ie, $p(r)$, collapsed across time, for $N=50$, and different values of $\epsilon_1$. The blue and red curves denote completely asynchronous and synchronous initial conditions, respectively, whereas the vertical line denotes $r_s$. \textcolor{red}{For all figures, $(\omega_0,\epsilon_2,\gamma)=(1.5,5.0,0.2)$}.}
    \label{N50_distribution}
\end{figure}

Real bio-physical systems, such as the brain, are typically modelled as a connected system with a finite number of constituent dynamical units (here, oscillators), the size of which can vary by different orders depending on the level of abstraction. Thus, while the previous sections demonstrate steady-state results for an infinitely large number of oscillators, the synchronisation phenomena in finite-sized systems are typically affected by fluctuations which can be better characterised by constructing a probability density of the order parameter, denoted by $p(r)$. \textcolor{red}{This is computed by simulating the system for a large number of time-steps (here, $10^5$) for a fixed set of system parameters. This is subsequently repeated over $100$ initial conditions, such that the first $50$ initial conditions correspond to $r(0)\approx 0.0$ (blue curve), and the remaining correspond to $r(0)\approx 1.0$ (red curve), with the results for $N=50$ shown in Fig.~\ref{N50_distribution}.}

It is observed that, for $\epsilon_1=0.6$ in Fig.~\ref{N50_distribution_a}, both distributions show a bimodal structure, with peaks near $r\approx 0.1$ and $r\approx 0.8$, respectively. However, the blue curve is substantially skewed towards $r=0$, whereas the red curve has two peaks of similar height. This indicates that, if initialised asynchronously, there is a higher probability of the system remaining asynchronous compared to transitioning to a highly synchronous state. On the other hand, if the system is initialised in a highly synchronous state, there is a nearly equal probability that the system will transition to an asynchronous state or remain in a highly synchronous state.

In comparison, for $\epsilon_1=1.3$ in Fig.~\ref{N50_distribution_b}, both distributions retain their bimodal structure. However, in contrast to the previous case, both distributions are skewed heavily towards $r=1$. Here, the red curve is much more sharply peaked near $r\approx 0.8$ compared to $r\approx 0.1$, indicating a higher probability of finding the system in a highly synchronous state if the system is initialised in a synchronous state. In comparison, the blue curve is only slightly skewed towards the synchronous state, indicating that the probability of finding the system in a synchronous state is marginally higher than that of finding it in an asynchronous state if the system is initialised in an asynchronous state. Finally, for $\epsilon_1=1.8$ and $2.6$, the distributions become nearly unimodal and skewed towards $r=1$, with a peak near $r\approx 0.8$. This implies that, for any initial condition, there is a much higher probability of observing the system to be in a highly synchronous state; see Figs.~\ref{N50_distribution_c} and \ref{N50_distribution_d}. \textcolor{red}{Taken together, the results presented in Fig.~\ref{N50_distribution} highlight the effects of size-induced fluctuations in driving the system to emergent states not observed in the infinite-size limit. Furthermore, additional plots of $p(r)$, obtained by repeating the same procedure for different values of $\epsilon_1$ and $N$, are presented in Fig.~\ref{N_distribution_transition} of \ref{app1}, which highlight the qualitative variation in the characteristic order parameter dynamics as a function of system size.}

\section{Conclusions \& Future Directions}
\label{conc}
\textcolor{red}{This work presents a very general theory of synchronization in a population of Kuramoto oscillators, where the interactions have both rotational symmetry-preserving and symmetry-breaking components. Furthermore, the additional terms are included such that they capture a wide range of coupling mechanisms, introduced through adaptation functions that modify the instantaneous coupling strength as a function of the synchronization order parameter, as well as the effects of frustration incorporated using a constant phase lag. Although these factors have been studied individually in the literature, this work introduces a unified mathematical framework to explore their collective interplay}. To this end, exact analytical results for the synchronisation states, their stability, and the corresponding phase transition points are obtained in the thermodynamic limit for any order of interaction \textcolor{red}{using the popular Ott-Antonsen Ansatz. Subsequently, these analytical results were verified through large-scale numerical simulations, focusing on the combined effects of pairwise and triadic interactions and an adaptation function having a quadratic dependence on the order parameter, revealing a repertoire of dynamical states, including monostable, bistable, and unstable synchronous cases}. Furthermore, varying the relative coupling strength between pairwise and higher-order interactions shows a qualitative change in the system dynamics, \textcolor{red}{highlighting both continuous and explosive phase transitions}, as well as a scenario where the desynchronization transition is suppressed. Finally, the effects of finite system size were also examined through numerically constructed order parameter distributions. The obtained results highlight how fluctuations in the temporal variation of the order parameter, arising out of the finite size of the system, can lead to transitions to a highly synchronous state that would not otherwise occur in the thermodynamic limit. \textcolor{red}{Such transitions occur spontaneously, \ie, without any change in system parameters, and are distinct from those observed in a system of nearly-infinite size, where transitions occur only upon tuning system parameters across a critical transition point.}

The results presented provide important insights into the interplay of higher-order interactions and other coupling modalities, allowing for more precise delineation of transition boundaries, yielding a comprehensive ``map'' of the accessible states. \textcolor{red}{Such models find widespread application across domains, among which a notable application lies in understanding epileptic seizures}, where incorporating higher-order interactions of different orders could improve surgical strategies for drug-resistant epilepsy~\cite{schramm2008temporal,jirsa2023personalised,wang2023delineating} by revealing underlying epileptogenic foci that can remain obscured in purely pairwise models, thereby facilitating precise and targeted interventions.  Another important example concerns the dissemination of information in social systems, where analysing higher-order connectivity can uncover underlying ``cells'', \ie, individuals or organisations that act as synchronised sources of misinformation, and could explain why misleading content often propagates far more rapidly than verified information~\cite{vosoughi2018spread}. 

\textcolor{red}{However, in spite of the theoretical advancements presented in this study, certain key limitations still remain. This includes obtaining the exact order of interaction driving the explosive spread of synchronous activity in various physical systems, as well as the explicit functional form of the adaptation functions governing the overall order parameter dynamics, thereby opening up many potential directions of future research. Furthermore, this will also include extending this framework to involve empirical hyper-graph connectivities (for example, see Ref.~\cite{ko2022growth}), integrating domain-specific models beyond the generic Kuramoto model, and estimating the various systems parameters using available datasets to obtain problem-specific insights}.

\section*{Author Contributions}
{\bf DB}: Conceptualization (equal);  Formal analysis (equal); Investigation (lead); Methodology (equal);  Software (lead); Validation (equal); Visualization (lead); Writing – original draft (equal).
{\bf AB}: Conceptualization (equal); Funding acquisition (lead); Investigation (supporting); Methodology (supporting); Project administration (lead); Supervision (lead); Visualization (supporting); Writing – review \& editing (lead).

\section*{Acknowledgements}
The authors acknowledge the financial support provided via NBRC Core Funds. DB acknowledges discussions with Dr. Proloy Das (NBRC), PV Vinitha (NBRC), and Shashwati Tripathi (NBRC).

\section*{Declaration of competing interest}
The authors declare that they have no known competing financial interests or personal relationships that could have appeared to influence the work reported in this paper. 

\section*{Data Availability Statement}
The data used in this article is available online~\cite{karmelic2022emergent,bougou2025mesoscale,dataset}.

\appendix
\section{Expression of Jacobians}
\label{app-jaco}

The stability of the steady states, denoted by $(r_s,\psi_s)$, of Eqs.~\eqref{reqn} and \eqref{psieqn} can be determined from the eigenvalues of their Jacobian $J$, the elements $J_{ij}$ of which are given by
\begin{align}
    \begin{split}
    J_{11}={}&\frac{-2r_s^2\Delta}{(1-r_s^2)}+r_s(1-r_s^2)\left[\sum_{d=1}^D\left.\frac{\D \sigma_d(r)}{\D r}\right|_{r_s}\cos(\delta_d)\right.+\\
    &\left.\sum_{d=2}^D\left.\frac{\D \sigma_d^{\prime}(r)}{\D r}\right|_{r_s}\cos(\chi_{d,s}^{\prime})\right],\label{J11}
    \end{split}\\
    \begin{split}
    J_{21}={}&\frac{-2r_s\omega_0}{(1+r_s^2)}-
    (1+r_s^2)\left[\sum_{d=1}^D\left.\frac{\D \sigma_d(r)}{\D r}\right|_{r_s}\sin(\delta_d)\right.+\\
    &\left.\sum_{d=2}^D\left.\frac{\D \sigma_d^{\prime}(r)}{\D r}\right|_{r_s}\sin(\chi_{d,s}^{\prime})\right],\label{J21}
    \end{split}\\
    J_{12}={}&2(h-1)r_s(1-r_s^2)\sum_{d=2}^D\sigma_d^{\prime}(r_s)\sin(\chi^{\prime}_{d,s}),\label{J12}\\
    \intertext{and}
    J_{22}={}&2(h-1)(1+r_s^2)\sum_{d=2}^D\sigma_d^{\prime}(r_s)\cos(\chi_{d,s}^{\prime}).\label{J22}
\end{align}
For the choice of parameter values of the simplified model in Sec. \ref{num-res}, the expressions for the elements of the Jacobian reduce to the form
\begin{align}
    \begin{split}
    J_{11}={}&\frac{-2r_s^2\Delta}{(1-r_s^2)}+r_s(1-r_s^2)\left[\left.\frac{\D \sigma_1(r)}{\D r}\right|_{r_s}+\right. \\ &\left. \frac{1}{\sigma_2(r_s)}\left(\frac{\Delta}{1-r_s^2}-\sigma_1(r_s)\right)\left.\frac{\D \sigma_2(r)}{\D r}\right|_{r_s}\right],
    \end{split}\label{j11_simple}\\
    J_{21}={}&\frac{-\omega_0}{(1+r_s^2)}\left[2r_s + \frac{1+r_s^2}{\sigma_2(r_s)}\left.\frac{\D \sigma_2(r)}{\D r}\right|_{r_s}\right],\\
    J_{12}={}&2r_s(1-r_s^2)\left[\frac{\omega_0}{1+r_s^2}\right],\\
    \intertext{and}
    J_{22}={}&2(1+r_s^2)\left[\frac{\Delta}{1-r_s^2}-\Big(\sigma_1(r_s)+\sigma_2(r_s)\Big)\right]\label{j22_simple},
\end{align}
where
\begin{equation}
    \sigma_1(r)=\frac{\epsilon_1}{2\left(1+\gamma r\right)^2},\quad \sigma_2(r)=\frac{\epsilon_2}{2}(1+\gamma r)^2 r^2,
\end{equation}
and
\begin{equation}
    \frac{\D \sigma_1(r)}{\D r}=\frac{-\gamma \epsilon_1}{(1+\gamma r)^3},\quad 
    \frac{\D \sigma_2(r)}{\D r}=\epsilon_2 r \left[1+\gamma r(3+2\gamma r)\right],
\end{equation}
respectively. Note that the expressions depend solely on $r_s$ in this case, thus eliminating the need to separately calculate $\psi_s$, thereby simplifying calculations.

\section{Simulation details}
\label{app2}

A direct simulation of the Kuramoto model, such as those with repeated initial conditions and different parameter values (\ie, Figs.~\ref{transition_diag} and \ref{N50_distribution}), with a coupling term of the form
\begin{equation}
    I=\sum_{\mathclap{j_1,..,j_d=1}}^N\sin\left(\sum_{k=1}^d c_k \phi_{j_k}+\sigma \phi_i + \alpha\right)
\label{gen-coup}
\end{equation}
would be prohibitively time-consuming due to the presence of nested summations, especially for $d\geq 2$. To address this, the RHS of Eq.~\eqref{gen-coup} is expanded as $I=X_d \cos(\sigma \phi_i+\alpha) + Y_d \sin(\sigma \phi_i+\alpha)$, where
\begin{equation}
    X_d=\sum_{\mathclap{j_1,..,j_d=1}}^N \sin\left(\sum_{k=1}^d c_k \phi_{j_k}\right),\quad
    Y_d=\sum_{\mathclap{j_1,..,j_d=1}}^N \cos\left(\sum_{k=1}^d c_k \phi_{j_k}\right).\label{tdp}
\end{equation}
Therefore, $X_d$ can be expanded as
\begin{multline}
    X_d=\sum_{j_1,..,j_d=1}^N\left[ \sin\left(\sum_{k=1}^{d-1} c_k \phi_{j_k} \right)\cos(c_d\phi_{j_d}) +\right. \\ \left. \cos\left(\sum_{k=1}^{d-1} c_k \phi_{j_k} \right)\sin(c_d\phi_{j_d})\right],
\end{multline}
which implies $X_d= X_{d-1} C_d + Y_{d-1} S_d$ and
\begin{equation}
    C_d=\sum_{j=1}^N \cos(c_d \phi_{j}),\quad  S_d=\sum_{j=1}^N \sin(c_d \phi_{j}).
\label{defs-cd-sd}
\end{equation}
Similarly, $Y_d=Y_{d-1} C_d - X_{d-1} S_d$, whereas $(X_1,Y_1)=(S_1,C_1)$. This provides an iterative method to efficiently compute the interaction terms of the form of Eq.~\eqref{gen-coup}, \textcolor{red}{and can be extended to Kuramoto-like coupling terms in any dynamical systems}.

\textcolor{red}{Unless mentioned otherwise, the simulations of Eq.~\eqref{pairwise_symmbreak} were performed using the popular 4th-order Runge-Kutta algorithm~\cite{butcher1996history}, for a system of $N=10^4$ oscillators for $10^4$ time-steps, with each individual time-step being $10^{-2}$ seconds.} The algorithm was implemented in C++, using OpenMP for parallelisation of the computationally intensive portions~\cite{dagum1998openmp}. The solutions of Eq.~\eqref{final-r-eqn} were estimated using the function fsolve(), whereas eigenvalues of the Jacobians in \ref{app-jaco} were obtained using the function eig() in MATLAB. All the computations were executed on a workstation equipped with an Intel Core i9-14900K processor and 64 GB DDR5 RAM.

\section{Varying system size}
\label{app1}

\begin{figure}[h]
    \centering
    \subcaptionbox{\label{N_distribution_transition_a}}{\includegraphics[scale=\sizeFig]{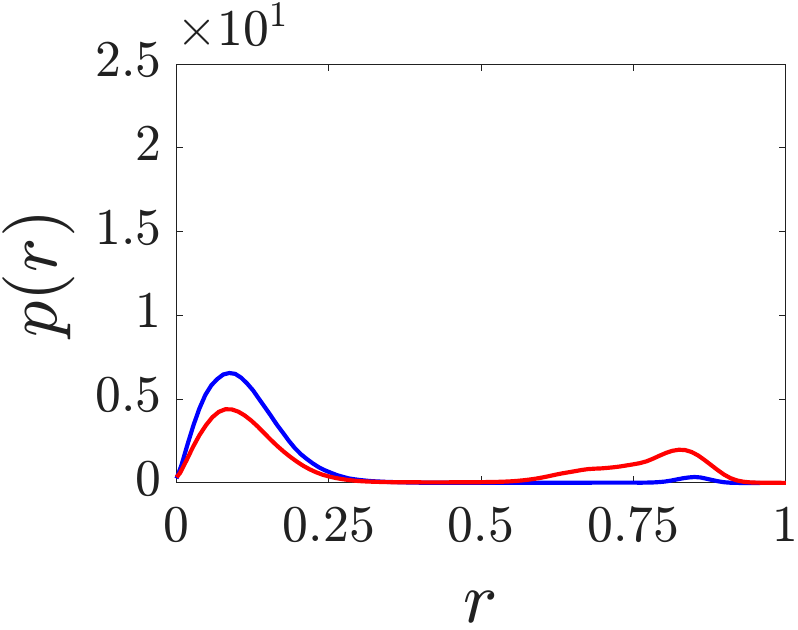}}
    \hfill
    \subcaptionbox{\label{N_distribution_transition_c}}{\includegraphics[scale=\sizeFig]{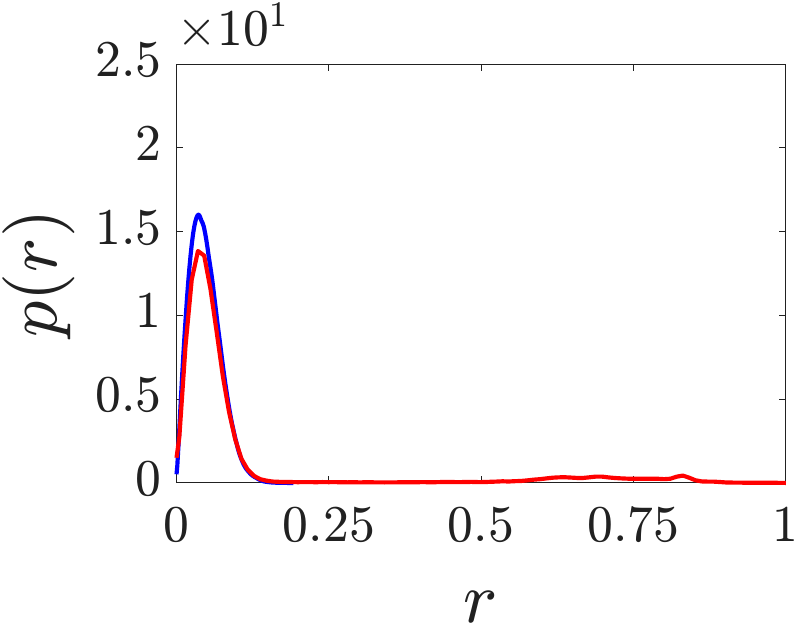}}
    \hfill
    \subcaptionbox{\label{N_distribution_transition_e}}{\includegraphics[scale=\sizeFig]{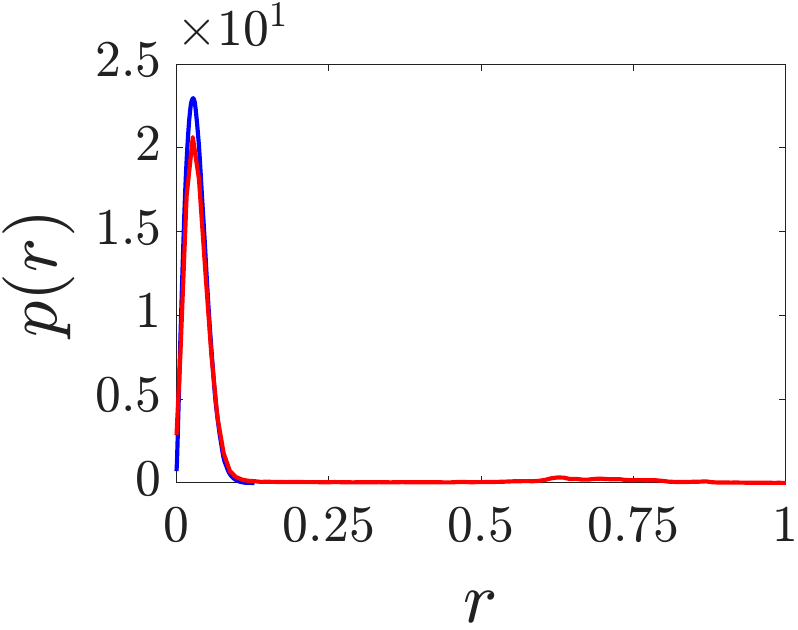}}

    \subcaptionbox{\label{N_distribution_transition_f}}{\includegraphics[scale=\sizeFig]{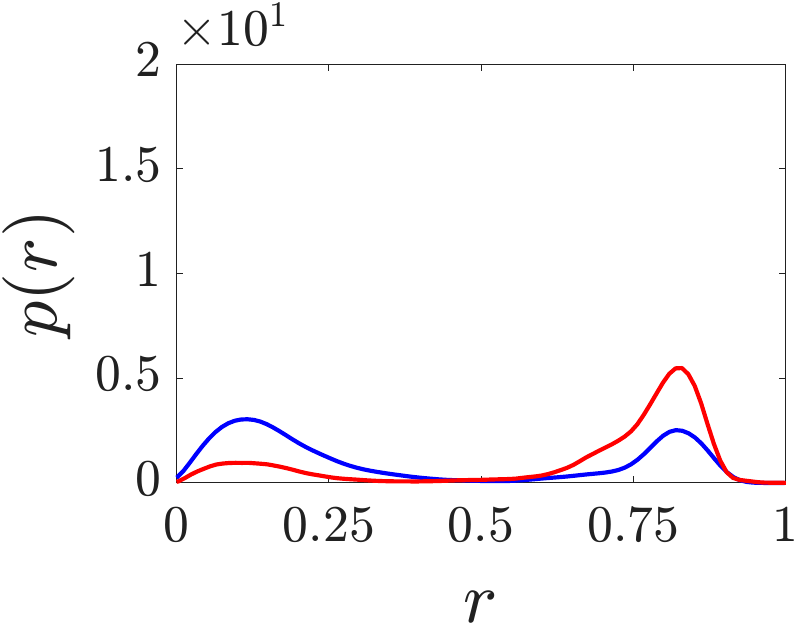}}
    \hfill
    \subcaptionbox{\label{N_distribution_transition_h}}{\includegraphics[scale=\sizeFig]{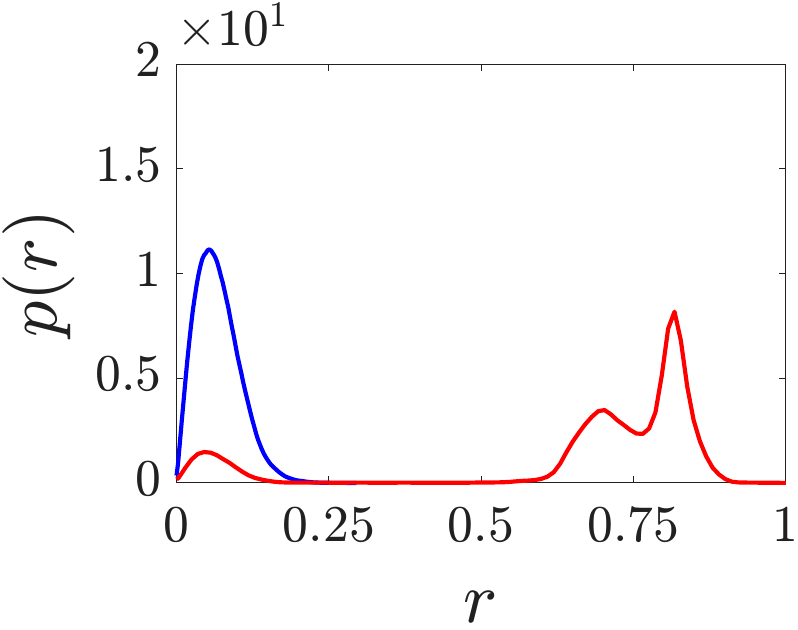}}
    \hfill
    \subcaptionbox{\label{N_distribution_transition_j}}{\includegraphics[scale=\sizeFig]{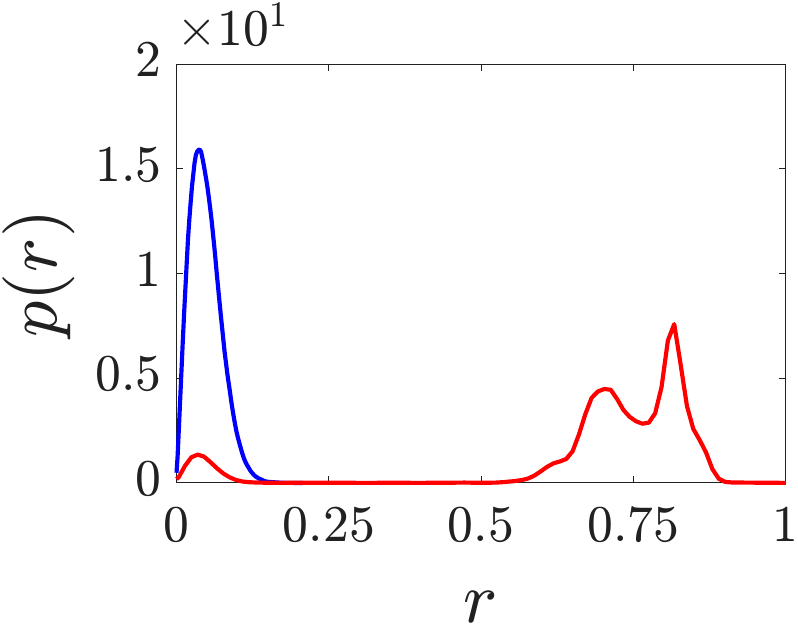}}

    \subcaptionbox{\label{N_distribution_transition_k}}{\includegraphics[scale=\sizeFig]{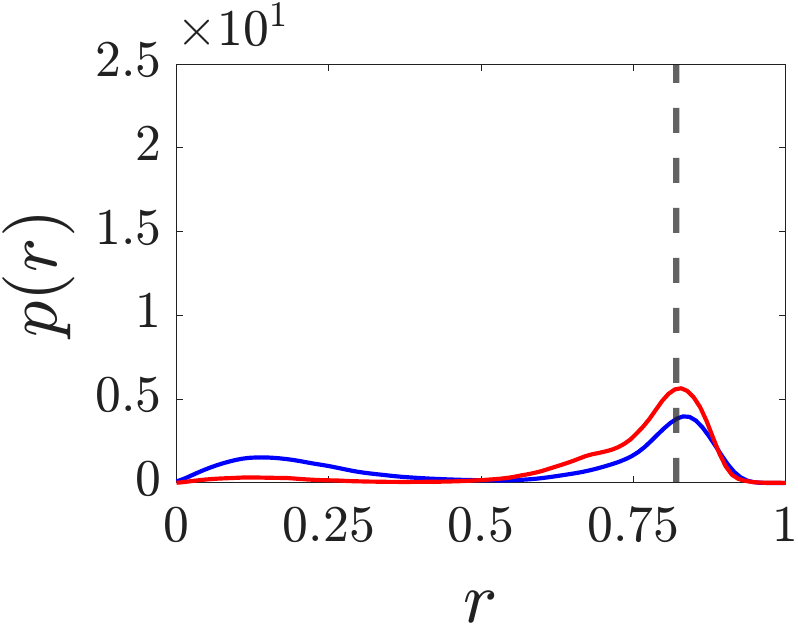}}
    \hfill
    \subcaptionbox{\label{N_distribution_transition_m}}{\includegraphics[scale=\sizeFig]{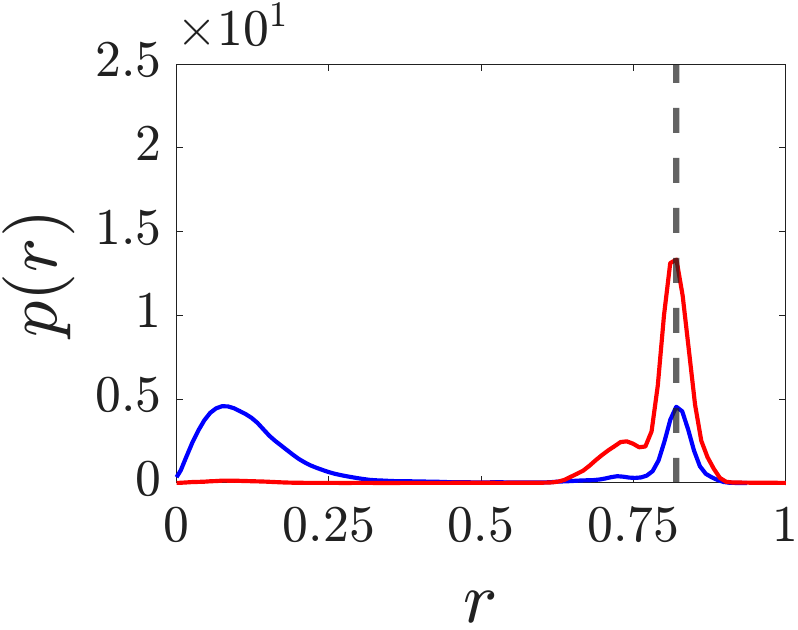}}
    \hfill
    \subcaptionbox{\label{N_distribution_transition_o}}{\includegraphics[scale=\sizeFig]{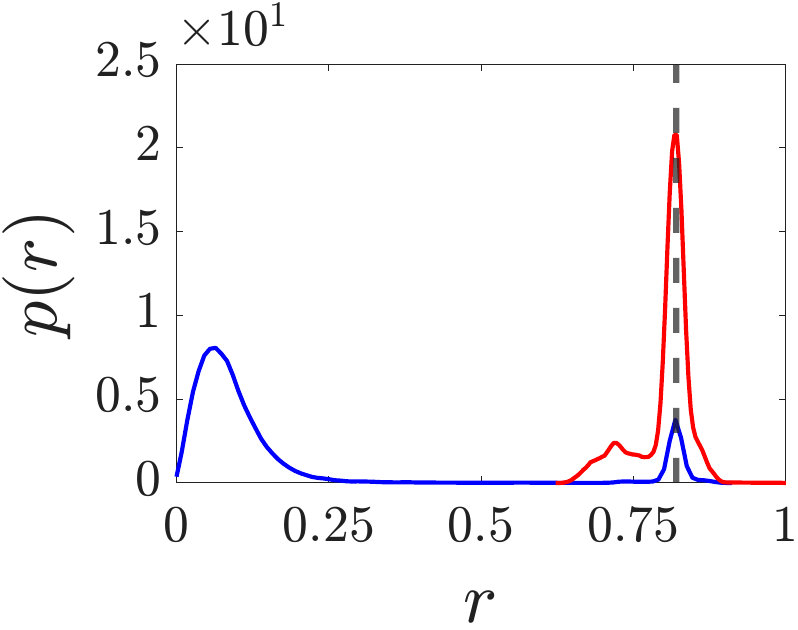}}

    \subcaptionbox{\label{N_distribution_transition_p}}{\includegraphics[scale=\sizeFig]{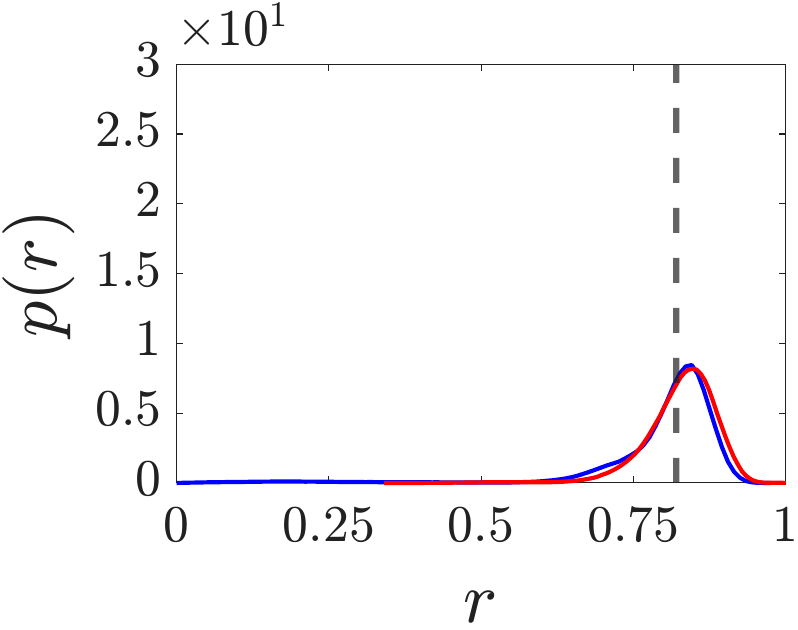}}
    \hfill
    \subcaptionbox{\label{N_distribution_transition_r}}{\includegraphics[scale=\sizeFig]{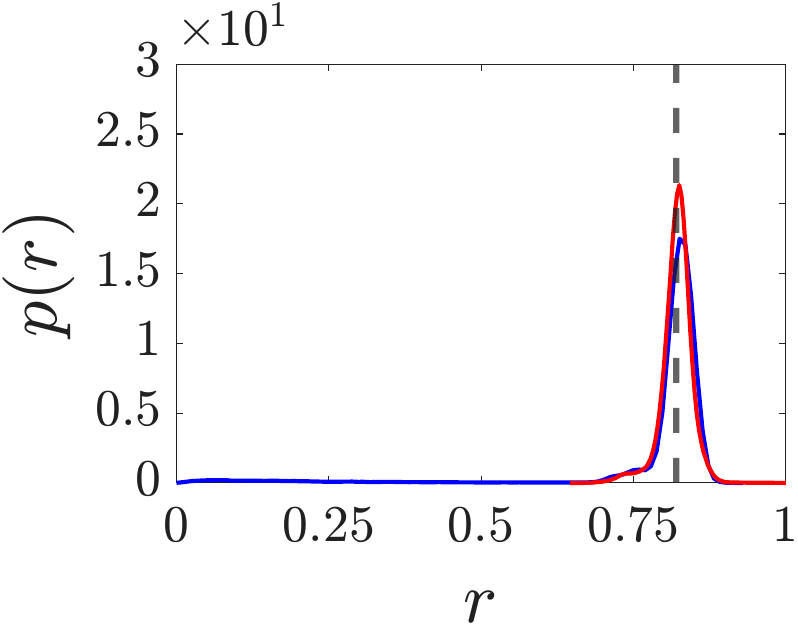}}
    \hfill
    \subcaptionbox{\label{N_distribution_transition_t}}{\includegraphics[scale=\sizeFig]{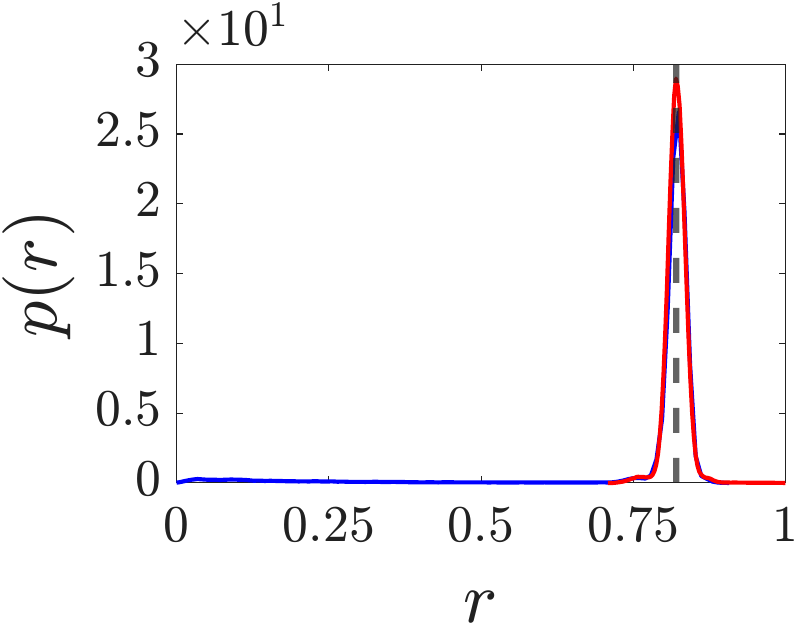}}

    \caption{(a)-(l): $p(r)$ for different values of $(N,\epsilon_1)$, where $N=100$, $500$, $1000$ (left to right), and $\epsilon_1=0.6$, $1.3$, $1.8$, $2.6$ (top to bottom). Here, $(\omega_0,\epsilon_2,\gamma)=(1.5,5.0,0.2)$, and the colours carry the same meaning as Fig.~\ref{N50_distribution}.}
    \label{N_distribution_transition}
\end{figure}

In Fig.~\ref{N_distribution_transition}, the figures highlight the characteristic change in the order parameter dynamics as the system size $N$ is increased for different values of the pairwise coupling strength $\epsilon_1$.
The results demonstrate that, for values of $\epsilon_1$ given by $0.6$, $1.3$, and $1.8$, the blue curve (asynchronous initial conditions) loses its bi-modal structure and becomes sharply peaked near $r=0$ as the value of $N$ is increased; see Fig.~\ref{N_distribution_transition_a}-\ref{N_distribution_transition_o}. In contrast, for $\epsilon_1=2.6$, the blue curve becomes sharply peaked near $r\approx 0.8$ upon increasing $N$; see Figs.~\ref{N_distribution_transition_p}-\ref{N_distribution_transition_t}. These observations are consistent with the fact that, in the limit $N\rightarrow\infty$, the system has a stable fixed point at $r=0$ $\forall$ $\epsilon_1\leq 2$. 
The observations regarding the red curve (synchronous initial conditions) are markedly different. In particular, for $\epsilon_1=0.6$, the red curve completely loses its bi-modal structure and becomes nearly identical to the blue curve with increasing $N$. This is consistent with previous analysis, which showed that $r=0$ is the only steady synchronous state for this parameter value in the case $N\rightarrow\infty$. However, upon increasing $N$ for $\epsilon_1=1.3$, the red curve becomes locally bi-modal around $r\approx 0.8$. This indicates the presence of oscillatory dynamics in the order parameter and is consistent with the previous findings. Finally, upon increasing $\epsilon_1$ to $1.8$ and $2.6$, the red curve becomes unimodal with a sharp peak near $r\approx 0.8$ as the value of $N$ is increased, which is indicative of a stable synchronous fixed point.

\bibliographystyle{unsrt}
\bibliography{refs}

@article{PhysRevE.111.014209,
  title = {Dynamics of swarmalators in the presence of a contrarian},
  author = {Sar, Gourab Kumar and Ansarinasab, Sheida and Nazarimehr, Fahimeh and Ghassemi, Farnaz and Jafari, Sajad and Ghosh, Dibakar},
  journal = {Phys. Rev. E},
  volume = {111},
  issue = {1},
  pages = {014209},
  numpages = {10},
  year = {2025},
  month = {Jan},
  publisher = {American Physical Society}
}

@article{salnikov2018simplicial,
  title={Simplicial complexes and complex systems},
  author={Salnikov, Vsevolod and Cassese, Daniele and Lambiotte, Renaud},
  journal={European Journal of Physics},
  volume={40},
  number={1},
  pages={014001},
  year={2018},
  publisher={IOP Publishing}
}

@article{lohe2015synchronization,
  title={Synchronization control in networks with uniform and distributed phase lag},
  author={Lohe, MA},
  journal={Automatica},
  volume={54},
  pages={114--123},
  year={2015},
  publisher={Elsevier}
}

@article{PhysRevE.96.042208,
  title = {{Synchronization scenarios in the Winfree model of coupled oscillators}},
  author = {Gallego, Rafael and Montbri\'o, Ernest and Paz\'o, Diego},
  journal = {Phys. Rev. E},
  volume = {96},
  issue = {4},
  pages = {042208},
  numpages = {11},
  year = {2017},
  month = {Oct},
  publisher = {American Physical Society}
}

@article{ansarinasab2025synchronization,
  title={{Synchronization transition in networks of Fractional-Order memristive Hindmarsh-Rose neurons}},
  author={Ansarinasab, Sheida and Nazarimehr, Fahimeh and Ghassemi, Farnaz and Jafari, Sajad},
  journal={Journal of Theoretical Biology},
  pages={112144},
  year={2025},
  publisher={Elsevier}
}

@article{ansarinasab2025transition,
  title={{Transition to synchronization in functional brain networks of children suffering from ADHD}},
  author={Ansarinasab, Sheida and Bayani, Atiyeh and Parastesh, Fatemeh and Ghassemi, Farnaz and Rajagopal, Karthikeyan and Jafari, Sajad},
  journal={Brain Structure and Function},
  volume={230},
  number={5},
  pages={82},
  year={2025},
  publisher={Springer}
}

@article{o2017oscillators,
  title={Oscillators that sync and swarm},
  author={O’Keeffe, Kevin P and Hong, Hyunsuk and Strogatz, Steven H},
  journal={Nature Communications},
  volume={8},
  number={1},
  pages={1504},
  year={2017},
  publisher={Nature Publishing Group UK London}
}

@article{ansarinasab2024spatial,
  title={The spatial dynamics and phase transitions in non-identical swarmalators},
  author={Ansarinasab, Sheida and Nazarimehr, Fahimeh and Sar, Gourab Kumar and Ghassemi, Farnaz and Ghosh, Dibakar and Jafari, Sajad and Perc, Matja{\v{z}}},
  journal={Nonlinear Dynamics},
  volume={112},
  number={12},
  pages={10465--10483},
  year={2024},
  publisher={Springer}
}

@article{PhysRevE.75.017201,
  title = {{Generalized coupling in the Kuramoto model}},
  author = {Filatrella, G. and Pedersen, N. F. and Wiesenfeld, K.},
  journal = {Phys. Rev. E},
  volume = {75},
  issue = {1},
  pages = {017201},
  numpages = {4},
  year = {2007},
  month = {Jan},
  publisher = {American Physical Society}
}

@article{buzanello2022matrix,
  title={{Matrix coupling and generalized frustration in Kuramoto oscillators}},
  author={Buzanello, Guilhermo L and Barioni, Ana Elisa D and de Aguiar, Marcus AM},
  journal={Chaos: An Interdisciplinary Journal of Nonlinear Science},
  volume={32},
  number={9},
  year={2022},
  publisher={AIP Publishing}
}

@article{PhysRevE.108.034208,
  title = {{Impact of phase lag on synchronization in frustrated Kuramoto model with higher-order interactions}},
  author = {Dutta, Sangita and Mondal, Abhijit and Kundu, Prosenjit and Khanra, Pitambar and Pal, Pinaki and Hens, Chittaranjan},
  journal = {Phys. Rev. E},
  volume = {108},
  issue = {3},
  pages = {034208},
  numpages = {7},
  year = {2023},
  month = {Sep},
  publisher = {American Physical Society}
}

@article{PhysRevE.102.012206,
  title = {Kuramoto model in the presence of additional interactions that break rotational symmetry},
  author = {Chandrasekar, V. K. and Manoranjani, M. and Gupta, Shamik},
  journal = {Phys. Rev. E},
  volume = {102},
  issue = {1},
  pages = {012206},
  numpages = {9},
  year = {2020},
  month = {Jul},
  publisher = {American Physical Society}
}

@article{mondal2025enhancing,
  title={Enhancing cluster synchronization in phase-lagged multilayer networks},
  author={Mondal, Abhijit and Khanra, Pitambar and Ghosh, Subrata and Kundu, Prosenjit and Hens, Chittaranjan and Pal, Pinaki},
  journal={Chaos, Solitons \& Fractals},
  volume={200},
  pages={117053},
  year={2025},
  publisher={Elsevier}
}

@article{PhysRevE.107.024311,
  title = {Population spiking and bursting in next-generation neural masses with spike-frequency adaptation},
  author = {Ferrara, Alberto and Angulo-Garcia, David and Torcini, Alessandro and Olmi, Simona},
  journal = {Phys. Rev. E},
  volume = {107},
  issue = {2},
  pages = {024311},
  numpages = {16},
  year = {2023},
  month = {Feb},
  publisher = {American Physical Society}
}

@article{ranjan2024propagation,
  title={Propagation of transient explosive synchronization in a mesoscale mouse brain network model of epilepsy},
  author={Ranjan, Avinash and Gandhi, Saurabh R},
  journal={Network Neuroscience},
  volume={8},
  number={3},
  pages={883--901},
  year={2024},
  publisher={MIT Press 255 Main Street, 9th Floor, Cambridge, Massachusetts 02142, USA}
}

@article{giusti2016two,
  title={{Two’s company, three (or more) is a simplex: Algebraic-topological tools for understanding higher-order structure in neural data}},
  author={Giusti, Chad and Ghrist, Robert and Bassett, Danielle S},
  journal={Journal of Computational Neuroscience},
  volume={41},
  number={1},
  pages={1--14},
  year={2016},
  publisher={Springer}
}

@article{lungeanu2021team,
  author = {Alina Lungeanu and Dorothy R. Carter and Leslie A. DeChurch and Noshir S. Contractor},
title = {{How Team Interlock Ecosystems Shape the Assembly of Scientific Teams: A Hypergraph Approach}},
journal = {Communication Methods and Measures},
volume = {12},
number = {2-3},
pages = {174--198},
year = {2018},
publisher = {Routledge},
doi = {10.1080/19312458.2018.1430756},
URL = {https://doi.org/10.1080/19312458.2018.1430756},
eprint = {https://doi.org/10.1080/19312458.2018.1430756}
}

@article{wang2020social,
  title={A social communication model based on simplicial complexes},
  author={Wang, Dong and Zhao, Yi and Leng, Hui and Small, Michael},
  journal={Physics Letters A},
  volume={384},
  number={35},
  pages={126895},
  year={2020},
  publisher={Elsevier}
}

@article{ghahremani2023novel,
  title={{A novel simplicial complex representation of social media texts: The case of Twitter}},
  author={Ghahremani, Yasamin and Amiri, Babak},
  journal={Chaos, Solitons \& Fractals},
  volume={173},
  pages={113642},
  year={2023},
  publisher={Elsevier}
}

@inproceedings{moore2012analyzing,
  title={{Analyzing collaboration networks using simplicial complexes: A case study}},
  author={Moore, Terrence J and Drost, Robert J and Basu, Prithwish and Ramanathan, Ram and Swami, Ananthram},
  booktitle={2012 Proceedings IEEE INFOCOM Workshops},
  pages={238--243},
  year={2012},
  organization={IEEE}
}

@article{bairey2016high,
  title={High-order species interactions shape ecosystem diversity},
  author={Bairey, Eyal and Kelsic, Eric D and Kishony, Roy},
  journal={Nature Communications},
  volume={7},
  number={1},
  pages={12285},
  year={2016},
  publisher={Nature Publishing Group UK London}
}

@article{ghasemi2022higher,
  title={Higher-order interaction learning of line failure cascading in power networks},
  author={Ghasemi, Abdorasoul and Kantz, Holger},
  journal={Chaos: An Interdisciplinary Journal of Nonlinear Science},
  volume={32},
  number={7},
  year={2022},
  publisher={AIP Publishing}
}

@article{gallardo2024higher,
  title={Higher-order interactions and emergent properties of microbial communities: The power of synthetic ecology},
  author={Gallardo-Navarro, Oscar and Aguilar-Salinas, Bernardo and Rocha, Jorge and Olmedo-{\'A}lvarez, Gabriela},
  journal={Heliyon},
  volume={10},
  number={14},
  year={2024},
  publisher={Elsevier}
}

@article{swain2022higher,
  title={Higher-order effects, continuous species interactions, and trait evolution shape microbial spatial dynamics},
  author={Swain, Anshuman and Fussell, Levi and Fagan, William F},
  journal={Proceedings of the National Academy of Sciences},
  volume={119},
  number={1},
  pages={e2020956119},
  year={2022},
  publisher={National Academy of Sciences}
}

@article{PhysRevE.109.024212,
  title = {{Prolonged hysteresis in the Kuramoto model with inertia and higher-order interactions}},
  author = {Sabhahit, Narayan G. and Khurd, Akanksha S. and Jalan, Sarika},
  journal = {Phys. Rev. E},
  volume = {109},
  issue = {2},
  pages = {024212},
  numpages = {11},
  year = {2024},
  month = {Feb},
  publisher = {American Physical Society}
}

@article{PhysRevE.101.032310,
  title = {Multibody interactions and nonlinear consensus dynamics on networked systems},
  author = {Neuh\"auser, Leonie and Mellor, Andrew and Lambiotte, Renaud},
  journal = {Phys. Rev. E},
  volume = {101},
  issue = {3},
  pages = {032310},
  numpages = {11},
  year = {2020},
  month = {Mar},
  publisher = {American Physical Society}
}

@article{PhysRevE.104.064305,
  title = {Consensus dynamics on temporal hypergraphs},
  author = {Neuh\"auser, Leonie and Lambiotte, Renaud and Schaub, Michael T.},
  journal = {Phys. Rev. E},
  volume = {104},
  issue = {6},
  pages = {064305},
  numpages = {13},
  year = {2021},
  month = {Dec},
  publisher = {American Physical Society}
}

@article{PhysRevE.111.014309,
  title = {Modification speed alters stability of ecological higher-order interaction networks},
  author = {Van Giel, Thomas and Daly, Aisling J. and Baetens, Jan M. and De Baets, Bernard},
  journal = {Phys. Rev. E},
  volume = {111},
  issue = {1},
  pages = {014309},
  numpages = {9},
  year = {2025},
  month = {Jan},
  publisher = {American Physical Society}
}

@book{hypergraphtheoryBretto2013,
  title     = {{Hypergraph Theory: An Introduction}},
  author    = "Bretto, Alain",
  year      = "2013",
  publisher = "Springer Cham",
  address   = "Switzerland"
}

@article{PhysRevE.102.012219,
  title = {{Dynamics of the generalized Kuramoto model with nonlinear coupling: Bifurcation and stability}},
  author = {Zou, Wei and Wang, Jianwei},
  journal = {Phys. Rev. E},
  volume = {102},
  issue = {1},
  pages = {012219},
  numpages = {14},
  year = {2020},
  month = {Jul},
  publisher = {American Physical Society}
}

@article{jin2023synchronization,
  title={Synchronization dynamics of phase oscillators with generic adaptive coupling},
  author={Jin, Xin and Wu, Yong-Gang and L{\"u}, Hua-Ping and Xu, Can},
  journal={Communications in Theoretical Physics},
  volume={75},
  number={4},
  pages={045601},
  year={2023},
  publisher={IOP Publishing}
}

@article{millan2025spatio,
  title={Spatio-temporal activity patterns induced by triadic interactions in an in silico neural medium},
  author={Mill{\'a}n, Ana P and Sun, Hanlin and Torres, Joaqu{\'\i}n J},
  journal={Journal of Physics: Complexity},
  volume={6},
  number={1},
  pages={015017},
  year={2025},
  publisher={IOP Publishing}
}

@article{arenas2008synchronization,
  title={Synchronization in complex networks},
  author={Arenas, Alex and D{\'\i}az-Guilera, Albert and Kurths, Jurgen and Moreno, Yamir and Zhou, Changsong},
  journal={Physics Reports},
  volume={469},
  number={3},
  pages={93--153},
  year={2008},
  publisher={Elsevier}
}

@article{RevModPhys.77.137,
  title = {{The Kuramoto model: A simple paradigm for synchronization phenomena}},
  author = {Acebr\'on, Juan A. and Bonilla, L. L. and P\'erez Vicente, Conrad J. and Ritort, F\'elix and Spigler, Renato},
  journal = {Rev. Mod. Phys.},
  volume = {77},
  issue = {1},
  pages = {137--185},
  numpages = {0},
  year = {2005},
  month = {Apr},
  publisher = {American Physical Society},
  doi = {10.1103/RevModPhys.77.137},
  url = {https://link.aps.org/doi/10.1103/RevModPhys.77.137}
}

@article{gupta2014kuramoto,
  title={Kuramoto model of synchronization: equilibrium and nonequilibrium aspects},
  author={Gupta, Shamik and Campa, Alessandro and Ruffo, Stefano},
  journal={Journal of Statistical Mechanics: Theory and Experiment},
  volume={2014},
  number={8},
  pages={R08001},
  year={2014},
  publisher={IOP Publishing}
}

@inproceedings{chen2023critical,
  title={Critical Dynamics of Kuramoto Model on Erd{\"o}s--R{\'e}nyi Random Graphs},
  author={Chen, Hai},
  booktitle={International Conference on Nonlinear Dynamics and Applications},
  pages={185--194},
  year={2023},
  organization={Springer}
}

@article{peron2019onset,
  title={{Onset of synchronization of Kuramoto oscillators in scale-free networks}},
  author={Peron, Thomas and Messias F. de Resende, Bruno and Mata, Ang{\'e}lica S and Rodrigues, Francisco A and Moreno, Yamir},
  journal={Phys. Rev. E},
  volume={100},
  number={4},
  pages={042302},
  year={2019},
  publisher={APS}
}

@article{tong2018exponential,
  title={{Exponential synchronization and phase locking of a multilayer Kuramoto-oscillator system with a pacemaker}},
  author={Tong, Dongbing and Rao, Pengchun and Chen, Qiaoyu and Ogorzalek, Maciej J and Li, Xiang},
  journal={Neurocomputing},
  volume={308},
  pages={129--137},
  year={2018},
  publisher={Elsevier}
}

@article{jain2023composed,
  title={{Composed solutions of synchronized patterns in multiplex networks of Kuramoto oscillators}},
  author={Jain, Priya B and Nguyen, Tung T and Min{\'a}{\v{c}}, J{\'a}n and Muller, Lyle E and Budzinski, Roberto C},
  journal={Chaos: An Interdisciplinary Journal of Nonlinear Science},
  volume={33},
  number={10},
  year={2023},
  publisher={AIP Publishing}
}

@article{yeung1999time,
  title={{Time delay in the Kuramoto model of coupled oscillators}},
  author={Yeung, MK Stephen and Strogatz, Steven H},
  journal={Phys. Rev. Lett.},
  volume={82},
  number={3},
  pages={648},
  year={1999},
  publisher={APS}
}

@article{wu2018dynamics,
  title={{Dynamics of Kuramoto oscillators with time-delayed positive and negative couplings}},
  author={Wu, Hui and Dhamala, Mukesh},
  journal={Phys. Rev. E},
  volume={98},
  number={3},
  pages={032221},
  year={2018},
  publisher={APS}
}

@article{ross2021dynamics,
  title={{Dynamics of coupled Kuramoto oscillators with distributed delays}},
  author={Ross, Aleksandra and Kyrychko, SN and Blyuss, KB and Kyrychko, YN},
  journal={Chaos: An Interdisciplinary Journal of Nonlinear Science},
  volume={31},
  number={10},
  year={2021},
  publisher={AIP Publishing}
}

@article{budzinski2023analytical,
  title={Analytical prediction of specific spatiotemporal patterns in nonlinear oscillator networks with distance-dependent time delays},
  author={Budzinski, Roberto C and Nguyen, Tung T and Benigno, Gabriel B and {\D}o{\`a}n, Jacqueline and Min{\'a}{\v{c}}, J{\'a}n and Sejnowski, Terrence J and Muller, Lyle E},
  journal={Phys. Rev. Res.},
  volume={5},
  number={1},
  pages={013159},
  year={2023},
  publisher={APS}
}

@article{costa2024bifurcations,
  title={{Bifurcations in the Kuramoto model with external forcing and higher-order interactions}},
  author={Costa, Guilherme S and Novaes, Marcel and de Aguiar, Marcus AM},
  journal={Chaos: An Interdisciplinary Journal of Nonlinear Science},
  volume={34},
  number={12},
  year={2024},
  publisher={AIP Publishing}
}

@article{costa2025exact,
  title={{Exact solutions of the Kuramoto model with asymmetric higher order interactions of arbitrary order}},
  author={Costa, Guilherme S and Novaes, Marcel and de Aguiar, Marcus AM},
  journal={Chaos, Solitons \& Fractals},
  volume={195},
  pages={116243},
  year={2025},
  publisher={Elsevier}
}

@article{pathak2022biophysical,
  title={Biophysical mechanism underlying compensatory preservation of neural synchrony over the adult lifespan},
  author={Pathak, Anagh and Sharma, Vivek and Roy, Dipanjan and Banerjee, Arpan},
  journal={Communications Biology},
  volume={5},
  number={1},
  pages={567},
  year={2022},
  publisher={Nature Publishing Group UK London}
}

@article{lehnertz2009synchronization,
  title={Synchronization phenomena in human epileptic brain networks},
  author={Lehnertz, Klaus and Bialonski, Stephan and Horstmann, Marie-Therese and Krug, Dieter and Rothkegel, Alexander and Staniek, Matth{\"a}us and Wagner, Tobias},
  journal={Journal of Neuroscience Methods},
  volume={183},
  number={1},
  pages={42--48},
  year={2009},
  publisher={Elsevier}
}

@article{osterhage2007measuring,
  title={{Measuring synchronization in the epileptic brain: A comparison of different approaches}},
  author={Osterhage, Hannes and Mormann, Florian and Staniek, Matth{\"a}us and Lehnertz, Klaus},
  journal={International Journal of Bifurcation and Chaos},
  volume={17},
  number={10},
  pages={3539--3544},
  year={2007},
  publisher={World Scientific}
}

@article{biswas2024symmetry,
  title={Symmetry-breaking higher-order interactions in coupled phase oscillators},
  author={Biswas, Dhrubajyoti and Gupta, Sayan},
  journal={Chaos, Solitons \& Fractals},
  volume={181},
  pages={114721},
  year={2024},
  publisher={Elsevier}
}

@article{biswas2024effect,
  title={Effect of adaptation functions and multilayer topology on synchronization},
  author={Biswas, Dhrubajyoti and Gupta, Sayan},
  journal={Phys. Rev. E},
  volume={109},
  number={2},
  pages={024221},
  year={2024},
  publisher={APS}
}

@article{ott2008low,
  title={Low dimensional behavior of large systems of globally coupled oscillators},
  author={Ott, Edward and Antonsen, Thomas M},
  journal={Chaos: An Interdisciplinary Journal of Nonlinear Science},
  volume={18},
  number={3},
  year={2008},
  publisher={AIP Publishing}
}

@article{ott2009long,
  title={Long time evolution of phase oscillator systems},
  author={Ott, Edward and Antonsen, Thomas M},
  journal={Chaos: An Interdisciplinary Journal of Nonlinear Science},
  volume={19},
  number={2},
  year={2009},
  publisher={AIP Publishing}
}

@article{majhi2022dynamics,
  title={{Dynamics on higher-order networks: A review}},
  author={Majhi, Soumen and Perc, Matja{\v{z}} and Ghosh, Dibakar},
  journal={Journal of the Royal Society Interface},
  volume={19},
  number={188},
  pages={20220043},
  year={2022},
  publisher={The Royal Society}
}

@article{shi2022simplicial,
  title={Simplicial networks: A powerful tool for characterizing higher-order interactions},
  author={Shi, Dinghua and Chen, Guanrong},
  journal={National Science Review},
  volume={9},
  number={5},
  pages={nwac038},
  year={2022},
  publisher={Oxford University Press}
}

@article{xu2021spectrum,
  title={{Spectrum of extensive multiclusters in the Kuramoto model with higher-order interactions}},
  author={Xu, Can and Skardal, Per Sebastian},
  journal={Phys. Rev. Res.},
  volume={3},
  number={1},
  pages={013013},
  year={2021},
  publisher={APS}
}

@article{biswas2022mirroring,
  title={{Mirroring of synchronization in a bi-layer master-slave configuration of Kuramoto oscillators}},
  author={Biswas, Dhrubajyoti and Gupta, Sayan},
  journal={Chaos: An Interdisciplinary Journal of Nonlinear Science},
  volume={32},
  number={9},
  year={2022},
  publisher={AIP Publishing}
}

@article{guo2021overviews,
  title={{Overviews on the applications of the Kuramoto model in modern power system analysis}},
  author={Guo, Yufeng and Zhang, Dongrui and Li, Zhuchun and Wang, Qi and Yu, Daren},
  journal={International Journal of Electrical Power \& Energy Systems},
  volume={129},
  pages={106804},
  year={2021},
  publisher={Elsevier}
}

@article{wang2021dynamic,
  title={{Dynamic synchronization of extreme heat in complex climate networks in the contiguous United States}},
  author={Wang, Zhi-Hua and Wang, Chenghao and Yang, Xueli},
  journal={Urban Climate},
  volume={38},
  pages={100909},
  year={2021},
  publisher={Elsevier}
}

@article{blanter2016kuramoto,
  title={Kuramoto model with non-symmetric coupling reconstructs variations of the solar-cycle period},
  author={Blanter, E and Le Mou{\"e}l, J-L and Shnirman, M and Courtillot, V},
  journal={Solar Physics},
  volume={291},
  number={3},
  pages={1003--1023},
  year={2016},
  publisher={Springer}
}

@inproceedings{fioriti2008stability,
  title={{Stability of a distributed generation network using the Kuramoto models}},
  author={Fioriti, Vincenzo and Ruzzante, Silvia and Castorini, Elisa and Marchei, Elena and Rosato, Vittorio},
  booktitle={International Workshop on Critical Information Infrastructures Security},
  pages={14--23},
  year={2008},
  organization={Springer}
}

@article{skardal2020higher,
  title={Higher order interactions in complex networks of phase oscillators promote abrupt synchronization switching},
  author={Skardal, Per Sebastian and Arenas, Alex},
  journal={Communications Physics},
  volume={3},
  number={1},
  pages={218},
  year={2020},
  publisher={Nature Publishing Group UK London}
}

@article{carballosa2023cluster,
  title={{Cluster states and $\pi$-transition in the Kuramoto model with higher order interactions}},
  author={Carballosa, Alejandro and Mu{\~n}uzuri, Alberto P and Boccaletti, Stefano and Torcini, Alessandro and Olmi, Simona},
  journal={Chaos, Solitons \& Fractals},
  volume={177},
  pages={114197},
  year={2023},
  publisher={Elsevier}
}

@article{sabhahit2024prolonged,
  title={{Prolonged hysteresis in the Kuramoto model with inertia and higher-order interactions}},
  author={Sabhahit, Narayan G and Khurd, Akanksha S and Jalan, Sarika},
  journal={Phys. Rev. E},
  volume={109},
  number={2},
  pages={024212},
  year={2024},
  publisher={APS}
}

@article{rajwani2023tiered,
  title={{Tiered synchronization in Kuramoto oscillators with adaptive higher-order interactions}},
  author={Rajwani, Priyanka and Suman, Ayushi and Jalan, Sarika},
  journal={Chaos: An Interdisciplinary Journal of Nonlinear Science},
  volume={33},
  number={6},
  year={2023},
  publisher={AIP Publishing}
}

@article{moyal2024rotating,
  title={{Rotating clusters in phase-lagged Kuramoto oscillators with higher-order interactions}},
  author={Moyal, Bhuwan and Rajwani, Priyanka and Dutta, Subhasanket and Jalan, Sarika},
  journal={Phys. Rev. E},
  volume={109},
  number={3},
  pages={034211},
  year={2024},
  publisher={APS}
}

@inproceedings{karmelic2022emergent,
  title={Emergent stability dynamics in the human brain connectome through the inclusion of high order interactions between coupled oscillators},
  author={Karmelic, Antonieta Michea and Perez-Acle, Tom{\'a}s and Ferreiro, M{\'o}nica Otero},
  booktitle={2022 41st International Conference of the Chilean Computer Science Society (SCCC)},
  pages={1--4},
  year={2022},
  organization={IEEE}
}

@article{manoranjani2023generalization,
  title={{Generalization of the Kuramoto model to the Winfree model by a symmetry breaking coupling}},
  author={Manoranjani, M and Gupta, Shamik and Senthilkumar, DV and Chandrasekar, VK},
  journal={The European Physical Journal Plus},
  volume={138},
  number={2},
  pages={144},
  year={2023},
  publisher={Springer}
}

@article{wen2023chemical,
  title={Chemical reaction networks and opportunities for machine learning},
  author={Wen, Mingjian and Spotte-Smith, Evan Walter Clark and Blau, Samuel M and McDermott, Matthew J and Krishnapriyan, Aditi S and Persson, Kristin A},
  journal={Nature Computational Science},
  volume={3},
  number={1},
  pages={12--24},
  year={2023},
  publisher={Nature Publishing Group US New York}
}

@article{brauch2013higher,
  title={Higher-order multicomponent reactions: beyond four reactants},
  author={Brauch, Sebastian and van Berkel, Sander S and Westermann, Bernhard},
  journal={Chemical Society Reviews},
  volume={42},
  number={12},
  pages={4948--4962},
  year={2013},
  publisher={Royal Society of Chemistry}
}

@article{sawicki2023perspectives,
  title={Perspectives on adaptive dynamical systems},
  author={Sawicki, Jakub and Berner, Rico and Loos, Sarah AM and Anvari, Mehrnaz and Bader, Rolf and Barfuss, Wolfram and Botta, Nicola and Brede, Nuria and Franovi{\'c}, Igor and Gauthier, Daniel J and others},
  journal={Chaos: An Interdisciplinary Journal of Nonlinear Science},
  volume={33},
  number={7},
  year={2023},
  publisher={AIP Publishing}
}

@article{skardal2022tiered,
  title={Tiered synchronization in coupled oscillator populations with interaction delays and higher-order interactions},
  author={Skardal, Per Sebastian and Xu, Can},
  journal={Chaos: An Interdisciplinary Journal of Nonlinear Science},
  volume={32},
  number={5},
  year={2022},
  publisher={AIP Publishing}
}

@article{yu2011higher,
  title={Higher-order interactions characterized in cortical activity},
  author={Yu, Shan and Yang, Hongdian and Nakahara, Hiroyuki and Santos, Gustavo S and Nikoli{\'c}, Danko and Plenz, Dietmar},
  journal={Journal of Neuroscience},
  volume={31},
  number={48},
  pages={17514--17526},
  year={2011},
  publisher={Soc Neuroscience}
}

@article{dagum1998openmp,
  title={{OpenMP: An Industry-Standard API for Shared-Memory Programming}},
  author={Dagum, Leonardo and Menon, Ramesh},
  journal={IEEE Computational Science and Engineering},
  volume={5},
  number={1},
  pages={46--55},
  year={1998},
  publisher={IEEE}
}

@article{butcher1996history,
  title={{A history of Runge-Kutta methods}},
  author={Butcher, John Charles},
  journal={Applied Numerical Mathematics},
  volume={20},
  number={3},
  pages={247--260},
  year={1996},
  publisher={Elsevier}
}

@article{boccaletti2023structure,
  title={The structure and dynamics of networks with higher order interactions},
  author={Boccaletti, Stefano and De Lellis, Pietro and Del Genio, CI and Alfaro-Bittner, Karin and Criado, Regino and Jalan, Sarika and Romance, Miguel},
  journal={Physics Reports},
  volume={1018},
  pages={1--64},
  year={2023},
  publisher={Elsevier}
}

@article{cabral2022metastable,
  title={Metastable oscillatory modes emerge from synchronization in the brain spacetime connectome},
  author={Cabral, Joana and Castaldo, Francesca and Vohryzek, Jakub and Litvak, Vladimir and Bick, Christian and Lambiotte, Renaud and Friston, Karl and Kringelbach, Morten L and Deco, Gustavo},
  journal={Communications Physics},
  volume={5},
  number={1},
  pages={184},
  year={2022},
  publisher={Nature Publishing Group UK London}
}

@article{bougou2025mesoscale,
  title={Mesoscale insights in Epileptic Networks: A Multimodal Intracranial Dataset},
  author={Bougou, Vasiliki and Vanhoyland, Micha{\"e}l and Cleeren, Evy and Janssen, Peter and Van Paesschen, Wim and Theys, Tom},
  journal={Scientific Data},
  volume={12},
  number={1},
  pages={774},
  year={2025},
  publisher={Nature Publishing Group UK London}
}

@misc{dataset,
  doi = {10.25493/ZC9M-2ES},
  url = {https://search.kg.ebrains.eu/instances/ffca40da-0e1e-42d8-bd17-ab04c7a5e4c9},
  author = {Bougou, Vasiliki and Vanhoyland, Michaël and Cleeren, Evy and Janssen, Peter and Van Paesschen, Wim and Theys, Tom},
  title = {{Mesoscale Insights in Epileptic Networks: A Multimodal Intracranial Dataset (v2)}},
  publisher = {EBRAINS},
  year = {2025},
  copyright = {The use of this dataset requires that the user cites the associated DOI and adheres to the conditions of use that are contained in the Data Use Agreement.}
}

@article{hindriks2025unraveling,
  title={{Unraveling high-order interactions in electrophysiological brain signals using elliptical distributions: moving beyond the Gaussian approximation}},
  author={Hindriks, Rikkert and van der Meulen, Frank and van Putten, Michel JAM and Tewarie, Prejaas},
  journal={Journal of Physics: Complexity},
  volume={6},
  number={2},
  pages={025011},
  year={2025},
  publisher={IOP Publishing}
}

@article{biswas2025effect,
  title={Effect of higher-order interactions on aging transitions of coupled neurons},
  author={Biswas, Dhrubajyoti and Seth, Soumyajit},
  journal={Chaos, Solitons \& Fractals},
  volume={200},
  pages={116977},
  year={2025},
  publisher={Elsevier}
}

@article{clusella2020irregular,
  title={{Irregular collective dynamics in a Kuramoto--Daido system}},
  author={Clusella, Pau and Politi, Antonio},
  journal={Journal of Physics: Complexity},
  volume={2},
  number={1},
  pages={014002},
  year={2020},
  publisher={IOP Publishing}
}

@article{PhysRevResearch.5.013074,
  title = {{Oscillation quenching in Stuart-Landau oscillators via dissimilar repulsive coupling}},
  author = {Dutta, Subhasanket and Alamoudi, Omar and Vakilna, Yash Shashank and Pati, Sandipan and Jalan, Sarika},
  journal = {Phys. Rev. Res.},
  volume = {5},
  issue = {1},
  pages = {013074},
  numpages = {12},
  year = {2023},
  month = {Feb},
  publisher = {American Physical Society},
  doi = {10.1103/PhysRevResearch.5.013074},
  url = {https://link.aps.org/doi/10.1103/PhysRevResearch.5.013074}
}

@article{odor2019critical,
  title={{Critical synchronization dynamics of the Kuramoto model on connectome and small world graphs}},
  author={{\'O}dor, G{\'e}za and Kelling, Jeffrey},
  journal={Scientific Reports},
  volume={9},
  number={1},
  pages={19621},
  year={2019},
  publisher={Nature Publishing Group UK London}
}

@article{heggli2019kuramoto,
  title={{A Kuramoto model of self-other integration across interpersonal synchronization strategies}},
  author={Heggli, Ole Adrian and Cabral, Joana and Konvalinka, Ivana and Vuust, Peter and Kringelbach, Morten L},
  journal={PLOS Computational Biology},
  volume={15},
  number={10},
  pages={e1007422},
  year={2019},
  publisher={Public Library of Science San Francisco, CA USA}
}

@article{PhysRevLett.120.244101,
  title = {{Kuramoto Model for Excitation-Inhibition-Based Oscillations}},
  author = {Montbri\'o, Ernest and Paz\'o, Diego},
  journal = {Phys. Rev. Lett.},
  volume = {120},
  issue = {24},
  pages = {244101},
  numpages = {6},
  year = {2018},
  month = {Jun},
  publisher = {American Physical Society},
  doi = {10.1103/PhysRevLett.120.244101},
  url = {https://link.aps.org/doi/10.1103/PhysRevLett.120.244101}
}

@article{odor2021effect,
  title={{The effect of noise on the synchronization dynamics of the Kuramoto model on a large human connectome graph}},
  author={{\'O}dor, G{\'e}za and Kelling, Jeffrey and Deco, Gustavo},
  journal={Neurocomputing},
  volume={461},
  pages={696--704},
  year={2021},
  publisher={Elsevier}
}

@article{PhysRevLett.131.207401,
  title = {Emergence and Control of Synchronization in Networks with Directed Many-Body Interactions},
  author = {Della Rossa, Fabio and Liuzza, Davide and Lo Iudice, Francesco and De Lellis, Pietro},
  journal = {Phys. Rev. Lett.},
  volume = {131},
  issue = {20},
  pages = {207401},
  numpages = {6},
  year = {2023},
  month = {Nov},
  publisher = {American Physical Society},
  doi = {10.1103/PhysRevLett.131.207401},
  url = {https://link.aps.org/doi/10.1103/PhysRevLett.131.207401}
}

@article{schramm2008temporal,
  title={{Temporal lobe epilepsy surgery and the quest for optimal extent of resection: A review}},
  author={Schramm, Johannes},
  journal={Epilepsia},
  volume={49},
  number={8},
  pages={1296--1307},
  year={2008},
  publisher={Wiley Online Library}
}

@article{jirsa2023personalised,
  title={Personalised virtual brain models in epilepsy},
  author={Jirsa, Viktor and Wang, Huifang and Triebkorn, Paul and Hashemi, Meysam and Jha, Jayant and Gonzalez-Martinez, Jorge and Guye, Maxime and Makhalova, Julia and Bartolomei, Fabrice},
  journal={The Lancet Neurology},
  volume={22},
  number={5},
  pages={443--454},
  year={2023},
  publisher={Elsevier}
}

@article{wang2023delineating,
  title={Delineating epileptogenic networks using brain imaging data and personalized modeling in drug-resistant epilepsy},
  author={Wang, Huifang E and Woodman, Marmaduke and Triebkorn, Paul and Lemarechal, Jean-Didier and Jha, Jayant and Dollomaja, Borana and Vattikonda, Anirudh Nihalani and Sip, Viktor and Medina Villalon, Samuel and Hashemi, Meysam and others},
  journal={Science Translational Medicine},
  volume={15},
  number={680},
  pages={eabp8982},
  year={2023},
  publisher={American Association for the Advancement of Science}
}

@article{vosoughi2018spread,
  title={The spread of true and false news online},
  author={Vosoughi, Soroush and Roy, Deb and Aral, Sinan},
  journal={Science},
  volume={359},
  number={6380},
  pages={1146--1151},
  year={2018},
  publisher={American Association for the Advancement of Science}
}

@article{ko2022growth,
  title={Growth patterns and models of real-world hypergraphs},
  author={Ko, Jihoon and Kook, Yunbum and Shin, Kijung},
  journal={Knowledge and Information Systems},
  volume={64},
  number={11},
  pages={2883--2920},
  year={2022},
  publisher={Springer}
}

@article{PhysRevE.111.044310,
  title = {{Brain wave dynamics in a Hopfield-Kuramoto model}},
  author = {Yao, Ruwei and Li, Yichao and Yao, Xintong and Wang, Kang and Qu, Jingling and Zou, Xiaolong and Hong, Bo},
  journal = {Phys. Rev. E},
  volume = {111},
  issue = {4},
  pages = {044310},
  numpages = {11},
  year = {2025},
  month = {Apr},
  publisher = {American Physical Society},
  doi = {10.1103/PhysRevE.111.044310},
  url = {https://link.aps.org/doi/10.1103/PhysRevE.111.044310}
}

\end{document}